\documentclass[aps, showpacs,superscriptaddress,nofootinbib]{revtex4-1}
\usepackage{ulem,graphicx,epsfig,amsmath,amsfonts,amssymb,xcolor,slashed}
\usepackage{lipsum}
\usepackage{epsfig}
\usepackage{epstopdf}
\usepackage{hyperref}
\usepackage{float}
\usepackage{wrapfig}
\usepackage[utf8]{inputenc}

\newcommand{\Slash}[1]{{\ooalign{\hfil/\hfil\crcr$#1$}}}

\renewcommand\sout{\bgroup \color[rgb]{1,0,0} \ULdepth=-.5ex \ULset}
\allowdisplaybreaks[1]
\begin{document}
\title{$\Lambda_b$ decays into 
 $\Lambda_c^*\ell\bar{\nu}_\ell$  and $\Lambda_c^*\pi^-$ $[\Lambda_c^*=\Lambda_c(2595)$ \& $\Lambda_c(2625)]$ and  heavy quark spin symmetry}
\date{\today}

\author{J.~Nieves}
\affiliation{Instituto~de~F\'{\i}sica~Corpuscular~(centro~mixto~CSIC-UV),
  Institutos~de~Investigaci\'on~de~Paterna, Aptdo.~22085,~46071,~Valencia,
  Spain}

\author{R.~Pavao}
\affiliation{Instituto~de~F\'{\i}sica~Corpuscular~(centro~mixto~CSIC-UV),
  Institutos~de~Investigaci\'on~de~Paterna, Aptdo.~22085,~46071,~Valencia,
  Spain}
\author{S.~Sakai}
\affiliation{Institute of Theoretical Physics, CAS, Zhong Guan Cun East Street 55 100190 Beijing, China}
\begin{abstract} 
We study the implications for  $\Lambda_b \to
  \Lambda_c^*\ell\bar{\nu}_\ell$ and  $\Lambda_b \to \Lambda_c^*\pi^-$ $[\Lambda_c^*=\Lambda_c(2595)$ and
    $\Lambda_c(2625)]$ decays that can be deduced from heavy quark spin symmetry (HQSS). Identifying the odd parity $\Lambda_c(2595)$ and $\Lambda_c(2625)$ resonances as HQSS partners, with  total angular momentum--parity $j_q^P=1^-$
  for the light degrees of freedom, we find that the ratios
  $\Gamma(\Lambda_b\rightarrow\Lambda_c(2595)\pi^-)/\Gamma(\Lambda_b\rightarrow\Lambda_c(2625)\pi^-)$
  and $\Gamma(\Lambda_b\rightarrow \Lambda_c(2595) \ell
  \bar{\nu}_\ell)/ \Gamma(\Lambda_b\rightarrow\Lambda_c(2625) \ell
  \bar{\nu}_\ell)$ agree, within errors, with the experimental values
  given in the Review of Particle Physics. We discuss how future, and more
  precise, measurements of the above branching fractions could be used
  to shed light into the inner HQSS structure of the narrow $\Lambda_c(2595)$
  odd-parity resonance. Namely, we show that such studies would constrain the
  existence of a sizable $j^P_q=0^-$ component in its wave-function, and/or of a two-pole
  pattern, in analogy to the case of the similar $\Lambda(1405)$
  resonance in the strange sector, as suggested by most of the
  approaches that describe the $\Lambda_c(2595)$ as a hadron molecule. We also investigate the lepton flavor universality ratios $R[\Lambda_c^*] = {\cal B}(\Lambda_b \to \Lambda_c^* \tau\,\bar\nu_\tau)/{\cal B}(\Lambda_b \to \Lambda_c^* \mu\,\bar\nu_\mu)$, and discuss how $R[\Lambda_c(2595)]$ may be affected by 
a new source of potentially large systematic errors if there are two  $\Lambda_c(2595)$ poles. 
\end{abstract}
\pacs{14.20.Lq, 14.40.Lb, 11.10.St, 12.38.Lg, 12.39.Hg, 13.30.-a}
\maketitle

\section{Introduction}
\label{sec_introduction}
Nowadays much attention is payed to the spectroscopy of heavy hadrons
in order to investigate the symmetries of Quantum Chromodynamics
(QCD). As pointed out in
Refs.~\cite{Isgur:1991wq,Wise:1992hn,Neubert:1993mb}, in the infinite
  quark mass limit ($m_Q\to \infty$), the spectrum of hadrons containing a heavy quark should show a
  SU(2)--pattern, because of the symmetry that QCD acquires in that
  limit under arbitrary rotations of the spin of the heavy quark. This
  is known as heavy quark spin symmetry (HQSS) in the literature.  In that case, the
  total angular momentum $j_q$ of the brown muck, which is the subsystem of the hadron apart
  from the heavy quark, is conserved and hadrons with $J=j_q\pm1/2$ form a degenerate doublet.  This
  is because the one gluon  exchange chromomagnetic interaction  between the heavy quark and the brown muck is suppressed
  by the infinitely large mass of the quark.

  Constituent quark models (CQMs) predict a nearly degenerate pair of
  $P-$wave $\Lambda_c^*$ excited states, with spin--parity $J^P=1/2^-$
  and $3/2^-$, whose masses are similar to those of the isoscalar
  odd-parity $\Lambda_c(2595)$ and $\Lambda_c(2625)$
  resonances~\cite{Copley:1979wj,Migura:2006ep,
    Garcilazo:2007eh,Roberts:2007ni,Yoshida:2015tia}. In the most
  recent of these CQM studies~\cite{Yoshida:2015tia}, two different
  types of excitation-modes are considered: The first one,
  $\lambda-$mode, accounts for excitations between the heavy quark and
  the brown muck as a whole, while the second one, $\rho-$mode,
  considers excitations inside the brown muck. When all quark masses
  are equal, $\lambda-$ and $\rho-$modes are
  degenerate~\cite{Yoshida:2015tia}. However for singly-heavy baryons,
  the typical excitation energies of the $\lambda-$mode are smaller
  than those of the $\rho-$mode. This is because for singly charm or
  bottom baryons, the interactions between the heavy quark and the
  brown muck are more suppressed than between the light
  quarks~\cite{Isgur:1991wr, Yoshida:2015tia}. Thus, one should expect
  the $\lambda$ excitation modes to become dominant for low-lying
  states of singly heavy-quark baryons. Within this picture, the
  $\Lambda^{\rm CQM}_c(2595)$ and $\Lambda^{\rm CQM}_c(2625)$ resonances would correspond
  to the members of the HQSS--doublet associated to
  $(\ell_\lambda=1,\ell_\rho=0)$, with total spin $S_q=0$ for the
  light degrees of freedom ({\it ldof}), leading to a  spin-flavor-spatial symmetric
  wave-function for the light isoscalar diquark subsystem inside of
  the $\Lambda_c^*$ baryon. The total spins of
  these states are the result of coupling the orbital-angular
  momentum $\ell_\lambda$ of the brown muck --with respect to the heavy
  quark-- with the spin ($S_Q$) of the latter. Thus both
  $\Lambda^{\rm CQM}_c(2595)$ and $\Lambda^{\rm CQM}_c(2625)$ states are connected by
  a simple rotation of the heavy-quark spin, and these resonances will be degenerate in
  the heavy-quark limit\footnote{The lowest-lying $\rho-$mode,
    $(\ell_\lambda=0,\ell_\rho=1)$ gives rise to two $\frac12^-$ and
    also two $\frac32^-$ multiplets of $\Lambda^*_c$'s, together with
    an additional $\frac52^-$ $\Lambda_c-$excited state, significantly
    higher in the spectrum~\cite{Yoshida:2015tia}. Note that the isoscalar
    light diquark could have $0^-$, $1^-$ and $2^-$ quantum-numbers,
    resulting from the coupling of the spin, $S_q=1$, and the
    orbital-angular momentum, $\ell_\rho=1$, of the light quarks. In the heavy quark limit all the baryons with the same light diquark $j_q^P$ configuration will be degenerate~\cite{Isgur:1991wr}.}.

  Since the total angular momentum and parity of the {\it ldof} in the $S-$wave $\pi\Sigma_c$ and $\pi\Sigma^*_c$ pairs are
  $1^-$, as in the CQM  $\Lambda_c(2595)$ and $\Lambda_c(2625)$ resonances, the $\Lambda^{\rm CQM}_c(2595) \to\pi\Sigma_c  \to  \pi\pi\Lambda_c$ and
  $\Lambda^{\rm CQM}_c(2625) \to \pi\Sigma^*_c  \to  \pi\pi\Lambda_c  $ decays
  respect HQSS, and hence   one should expect sizable widths for these
  resonances, unless these   transitions are  kinematically
  suppressed. This scenario seems plausible, as can be inferred
  from the masses and thresholds compiled in Table~\ref{tab:mass}. Indeed, the recent
  works of Refs.~\cite{Nagahiro:2016nsx, Arifi:2017sac} find widths
  for the CQM $(\ell_\lambda=1,\ell_\rho=0)$ states ($j_q^P=1^-$)  predicted in
  \cite{Yoshida:2015tia} consistent with data.
  \begin{table*}[b!]
    \begin{tabular}{c|ccccc}
      & $M$ & $\Gamma$ & $M(\Sigma_c^{(*)+}+\pi^0)$ &
      $M(\Sigma_c^{(*)0}+\pi^+)$ &  $M(\Sigma_c^{(*)++}+\pi^-)$
      \\ \hline
      $\Lambda_c(2595)$&$2592.25\pm 0.28$& $2.6\pm 0.6$ & $2587.9 \pm
      0.4$ & $2593.32 \pm 0.14$   & $2593.54 \pm 0.14$ \\
      $\Lambda_c(2625)$&$2628.11\pm 0.19$& $< 0.97$ & $2652.5 \pm
      2.3$ & $2658.05 \pm 0.20$   & $2657.98 \pm 0.20$  \\\hline
    \end{tabular}
    \caption{Masses and widths of the  $\Lambda_c(2595)$ and $\Lambda_c(2625)$
  resonances (MeV units). Thresholds (MeV) of some possible $S-$wave decay
  channels are also given. In addition, the thresholds of the three-body
  channels, after the $P-$wave decay of the $\Sigma_c^{(*)}$
  resonances, are $M(\Lambda_c+\pi^++\pi^-)= 2565.60 \pm 0.14$ MeV
     and  $M(\Lambda_c+\pi^0+\pi^0)= 2556.41 \pm 0.14$ MeV. Data taken
  from the   Review of Particle Physics (RPP)~\cite{Tanabashi:2018oca}.\label{tab:mass}}
  \end{table*}
 
A different mechanism to explain the small width of the  $\Lambda_c (2595) $ would be that 
its wave-function had a large $j_q^P=0^-$ {\it ldof} component\footnote{Note that, in principle, both $j_q=0^-$ and $j_q=1^-$
  configurations can couple with the spin ($S_Q=\frac12$) of the charm
  quark to give a total $J^P = \frac12^-$ for the $\Lambda_c(2595)$. }. This is because the transition of this $j_q^P=0^-$
term of the $\Lambda_c(2595)$ to the final $\pi\Sigma_c$ state
will be suppressed by HQSS. This new mechanism will act in addition to
any possible kinematical suppression. As we will see in the next section, it turns out that some of the
approaches that describe the $ \Lambda_c(2595)$ as a hadron-molecule
predict precisely a significant $j_q^P=0^-$ component for the inner HQSS
structure of this resonance. These models generate also the existence
of a second, broad, resonance in the region of the $\Lambda_c(2595)$,
with a large $j_q^P=1^-$ {\it ldof} component, that could be naturally identified
to the HQSS partner of the $\Lambda_c(2625)$, since both states will have
the same brown muck configuration in the heavy-quark limit\footnote{Since the spin-parity
  of the $\Lambda_c(2625)$ is $\frac32^-$ and  it is the lowest-lying
  state  with these quantum numbers, one should expect the total angular momentum and parity of the
  {\it  ldof} in the $\Lambda_c(2625)$ to be $1^-$.}.

In this work, we will derive HQSS relations between the $\Lambda_b$ decays into
$\Lambda_c^*\pi^-$ and $\Lambda_c^*\ell\bar{\nu}_\ell$
$[\Lambda_c^*=\Lambda_c(2595)$ and $\Lambda_c(2625)]$, supposing firstly that the $\Lambda_c(2595)$ and $\Lambda_c(2625)$ form the lowest-lying $j_q^P=1^-$ HQSS doublet. We will also discuss
how  measurements of the ratio of branching fractions  
$\Gamma[\Lambda_b\rightarrow \Lambda_c(2595)]/\Gamma[\Lambda_b\rightarrow \Lambda_c(2625)]$  can
be used to constrain the existence of a sizable $j^P_q=0^-$ {\it ldof} component
in the $\Lambda_c(2595)$ wave-function, and/or of a second pole,
in analogy to the case of the similar $\Lambda(1405)$ resonance.

Exclusive semileptonic $\Lambda_b$ decays into excited charmed
$\Lambda_c(2595)$ and $\Lambda_c(2625)$ baryons have been studied
using heavy quark effective theory (HQET), including order $\Lambda_{\rm QCD}
/m_Q$ corrections~\cite{Roberts:1992xm, Leibovich:1997az}, and  
non-relativistic and semi-relativistic CQMs~\cite{Pervin:2005ve}, always assuming a single pole structure
for the first of these resonances and a dominant $j^P_q=1^-$
configuration. Recently, it has also been 
suggested that  measurements of these decays by
LHCb could be used to perform precise lepton flavor universality (LFU) tests~\cite{Boer:2018vpx, Gutsche:2018nks},  comparing branching fractions with $\tau-$ or $\mu-$leptons in the
final state. 
The analyses of Refs.~\cite{Boer:2018vpx} and \cite{Gutsche:2018nks} assumed that both excited charmed baryons
form a doublet under HQSS, and therefore it neither
contemplated the possibility that the narrow $\Lambda_c(2595)$ might
not be the HQSS partner of the $\Lambda_c(2625)$, nor that it could contain
a non-negligible $j_q^P=0^-$ component, as it occurs in most of the
molecular descriptions of this resonance. It is therefore timely and of the utmost
interest to test the HQSS doublet assumption for the $\Lambda_c(2595)$ and $\Lambda_c(2625)$ with the available data.

A first step in that direction was given in Refs.~\cite{Liang:2016exm,
  Liang:2016ydj}. In these two works, the semileptonic $\Lambda_b \to \Lambda_c^*$ transitions, together with
the $\Lambda_b$ decays into $\Lambda_c^*\pi^-$ and $\Lambda_c^*D_s^-$ were studied. It
was found that the ratios of the rates obtained for $\Lambda_c(2595)$
and $\Lambda_c(2625)$ final states are very sensitive to the couplings
of these resonances to the $D^*N$ channel, which also becomes essential
to obtain agreement with the available data. Following the claims of
Refs.~\cite{Liang:2016exm, Liang:2016ydj}, these results seem to give
strong support to the molecular picture of the two $\Lambda_c^*$
states, and the important role of the $D^*N$ component in their dynamics\footnote{The
  same type of ideas were extended in Ref.~\cite{Pavao:2017cpt} to the
  semileptonic and one pion decays of the $\Xi_b^-$ baryons into
  $\Xi_c^*$ resonances, analogs of the $\Lambda_c(2595)$ and
  $\Lambda_c(2625)$ states in the charm-strange sector.}.  As we will
discuss in the next section, the $\Lambda_c(2595) D^*N$ and
$\Lambda_c(2625) D^*N$ couplings, together with those to the $DN$ and
$\pi\Sigma_c^{(*)}$ pairs, can also be used to obtain valuable information
on the inner HQSS structure of these resonances.

Within a manifest Lorentz and
HQSS invariant formalism~\cite{Georgi:1990cx,Mannel:1990vg,Falk:1991nq},  we will re-examine here some of the results obtained in
Refs.~\cite{Liang:2016exm, Liang:2016ydj}, and  will connect
the findings of these two works with the quantum numbers of the  {\it ldof} in the $\Lambda_c(2595)$ wave function. Specifically, 
we will  discuss how  future accurate measurements of the 
different ratios of branching fractions proposed in \cite{Liang:2016exm, Liang:2016ydj} may be used to constrain or
discard  i) a sizable $j^P_q=0^-$ component in the
$\Lambda_c(2595)$ wave-function, and ii) the existence of a second pole, analog  to the second (broad) $\Lambda(1405)$ resonance~\cite{Tanabashi:2018oca}. 
The study will also shed some light on the validity of some of the most popular hadron-molecular interpretations of the 
odd-parity lowest-lying $\Lambda_c^*$ states. 

This work is structured as follows. After this introduction, in Sec.~\ref{sec:mol} we critically review different 
molecular descriptions of  the $\Lambda_c(2595)$ and $\Lambda_c(2625)$ baryons, and  discuss in detail the main features of those models that predict a two-pole  pattern for the $\Lambda_c(2595)$. Next in  Sec.~\ref{sec:SL-decay}, we study the semileptonic $\Lambda_b \to \Lambda_c^*\ell\bar{\nu}_\ell$ decays and the constrains imposed by HQSS to these processes. We derive a scheme that preserves spin-symmetry in  the $b-$quark sector and that leads 
to simple and accurate expressions for the differential widths, including ${\cal O}(1/m_c)$ corrections and full finite-lepton mass contributions that
are necessary for testing LFU. Semileptonic decays to molecular $\Lambda_c^{\rm MOL}$ states are addressed in Subsec.~\ref{sec:molecular-decays}, and the pion mode is examined in  Sec.~\ref{sec:pidecay}. The numerical results of this work are presented in Sec.~\ref{sec:results}. First in 
Subsec.~\ref{sec:results-massless}, we discuss the semileptonic ($\mu^- \bar \nu_\mu$ or $e^- \bar \nu_e$) and pion $\Lambda_b\to \Lambda_c^*$ decays,  and  present $m_Q\to \infty$, ${\cal O}(1/m_Q)$ HQET and molecular-model predictions for the ratios of branching fractions studied in 
\cite{Liang:2016exm, Liang:2016ydj}. Next in Subsec.~\ref{sec:results-tau}, we show results for  $\Lambda_b$ semileptonic decays with a $\tau$ lepton in the final state that can be of interest for LFU tests. Finally, we outline the main conclusions of this work in Sec.~\ref{sec:summary}.

\section{HQSS structure of the $\Lambda_c(2595)$
  and $\Lambda_c(2625)$ states in hadron-molecular approaches.}\label{sec:mol}

In this section, we will discuss the most important common features and results obtained from
approaches where the $ \Lambda_c(2595)$ and $\Lambda_c (2625)$ are
described as hadron-molecules. These studies are 
motivated by the appealing similitude of these resonances to the
$\Lambda(1405) $ and $ \Lambda(1520) $ in the strange sector. In
particular the two isoscalar $S$-wave $\Lambda(1405)$ and
$\Lambda_c(2595)$ resonances have several features in common. The mass
of the former lies in between the $\pi\Sigma$ and $\bar K N$ channel
thresholds, to which it couples
strongly~\cite{Dalitz:1959dn,Dalitz:1960du,Dalitz:1967fp}.  In turn, the
$\Lambda_c(2595)$ lies below the $DN$ and just slightly above the
$\pi\Sigma_c$ thresholds, and substituting the $c$ quark by a $s$
quark, one might expect the interaction  of $DN$ to play a role in the dynamics of the $\Lambda_c(2595)$ similar to that played by 
$\bar{K}N$ in the strange sector.

The hadronic molecular interpretation of the $\Lambda(1405)$ provides
a good description of its properties. Actually, the dynamics of this
resonance is mostly governed by the leading order (LO) SU(3) chiral Weinberg-Tomozawa (WT)
meson-baryon interaction. The resonance is dynamically generated
 from the interaction of the mesons of the $0^-$ octet
(Goldstone bosons) with the $1/2^+$ octet of ground state
baryons~\cite{Kaiser:1995eg, Oset:1997it,Oller:2000fj, Lutz:2001yb,
  GarciaRecio:2002td,Hyodo:2002pk,Jido:2003cb, GarciaRecio:2003ks}
(see also the most recent works of Refs.~\cite{Hyodo:2011ur, Kamiya:2016jqc}
and references therein for further details and other related studies on
the $\Lambda(1405)$). One of the distinctive features of this
resonance is its two-pole structure~\cite{Oller:2000fj,GarciaRecio:2002td,Hyodo:2002pk,Jido:2003cb,GarciaRecio:2003ks,Hyodo:2011ur,
  Kamiya:2016jqc}, that have found experimental confirmation
\cite{Thomas:1973uh,Prakhov:2004an} as discussed in Ref.~\cite{Magas:2005vu}. This
two-pole pattern\footnote{One narrow state situated below the $\bar
  K N$ threshold and with a small coupling to the open $\pi\Sigma$
  channel, and a second state much wider because its large coupling to
  the open $\pi\Sigma$ channel.}  is by now widely accepted by the
community (see f.i. the mini review on this issue in the RPP
by the Particle Data Group~\cite{Tanabashi:2018oca}). 

On the other hand, many works have been also devoted to the study of dynamically generated $J^P=3/2^-$ states in the SU(3)
sector~\cite{Kolomeitsev:2003kt,Sarkar:2004jh,Sarkar:2005ap,Roca:2006sz,Hyodo:2006uw,Doring:2006ub,GarciaRecio:2005hy,Toki:2007ab,
  Sarkar:2009kx, Lutz:2001mi, Oset:2009vf, Gamermann:2011mq}. Early works considered
only the chiral interaction of pseudoscalar $0^-$ mesons with the baryons
of the $3/2^+$ decuplet, but more recently,  vector-mesons degrees of
freedom have also been incorporated in the coupled-channel approach,
using different schemes (see for instance the discussion in
\cite{Gamermann:2011mq}). In these approaches, the $\Lambda(1520)$ is
dynamically generated mostly from the $S-$wave $\pi\Sigma^*-\bar K^* N$
coupled-channels dynamics, appearing it slightly above the $\pi\Sigma^*$
threshold. It has a non-vanishing width, since  the $\pi\Sigma^*$ channel
is open. In clear analogy, one might naturally think of a similar
mechanism to generate the $\Lambda_c(2625)$ from the $\pi\Sigma^*_c-D^*
N$ dynamics, though the major difference is that the charm-resonance
is located around 30-25 MeV below the $\pi\Sigma^*_c$ threshold. 

\subsection{Molecular models}

The general scheme  consists of taking some $S-$wave interactions as kernel of a Bethe-Salpeter equation (BSE), conveniently ultraviolet (UV) renormalized,
and whose solutions fulfill exact elastic unitarity in
coupled-channels. In this context, bound and resonant states appear as
poles in the appropriate Riemann-sheets\footnote{This is in gross
  features also the scheme used in the previous works on the
  $\Lambda(1405)$ and $\Lambda(1520)$, and in most of the studies leading to
  hadron-molecular interpretations of many other resonances.}, and the residues provide the coupling of the dynamically generated states to the different channels considered in the approach.

The resemblance of the physics in the odd-parity charm $C=1$ baryon sector to the
phenomenology seen in $\bar K N-\pi \Sigma$ dynamics was first exploited in
the works of Refs.~\cite{Lutz:2003jw, Tolos:2004yg}.  These
first two works had some clear limitations. In the first one, the $J^P= 1/2^-$ sector is studied using  
the scattering of Goldstone bosons off $1/2^+$ heavy-light baryon resonances. Despite
the interactions were fully consistent with chiral symmetry,  neither the $DN$, nor the $D^*N$
channels were  considered~\cite{Lutz:2003jw}.  The work of Ref.~\cite{Tolos:2004yg} also studied the $\Lambda_c(2595)$ and there, the
interactions were obtained from chirally motivated Lagrangians upon
replacing the $s$ quark by the $c$ quark. Though in this way, the $DN$ channel was accounted for, the HQSS counterpart $D^*N$  was not considered.   

The subsequent works of Refs.~\cite{Hofmann:2005sw, Mizutani:2006vq} and \cite{Hofmann:2006qx} for the $J^P=3/2^-$ sector, introduced some
improvements on the schemes of Refs.~\cite{Lutz:2003jw, Tolos:2004yg}. Namely, the BSE interaction kernels  were obtained from $t$-channel
exchange of vector mesons between pseudoscalar mesons and baryons, in
such a way that chiral symmetry is preserved in the light meson
sector. Besides, the universal vector meson coupling hypothesis [Kawarabayashi-Suzuki-Fayyazuddin-Riazudden (KSFR)~\cite{Kawarabayashi:1966kd, Riazuddin:1966sw}] was modified to take into account the reduction of the interaction strength provoked by the mass of the $t-$channel exchanged meson. In this way, some SU(4) flavor-symmetry breaking corrections, additional to those induced by 
the use of the physical masses, were considered. Similar qualitative findings were obtained in the
work of Ref.~\cite{JimenezTejero:2009vq}, where some finite range effects were explored.

A detailed treatment of the interactions between the ground-state singly charmed and bottomed baryons and the pseudo-Nambu-Goldstone
bosons, discussing also  the effects of the next-to-leading-order chiral potentials, was carried out in \cite{Lu:2014ina}. However,  channels not involving  Goldstone bosons, like $DN$ or $D^*N$,  were again not considered.  In this reference, several aspects related to the renormalization procedure were also critically discussed\footnote{It is also worth mentioning  Ref.~\cite{Lu:2016gev}, where  the properties of the $\Lambda_c(2595)$  are discussed in the limit of large number of colors ($N_c$), within  several schemes. The $N_c\gg 3$ behaviour of the resonance properties (mass, width, couplings, etc.) puts constrains on its possible dynamical origin, since  the importance of the unitary loops involving Goldstone 
bosons decreases as $N_c$ grows~\cite{Pelaez:2006nj, GarciaRecio:2006wb, Nieves:2009ez, Nieves:2011gb}.}. 

In all cases, the $\Lambda_c(2595)$,  or  the $\Lambda_c(2625)$ if studied, could be dynamically generated after a
convenient tuning of the renormalization constants. However, none of
these works were consistent with HQSS since none of them considered the $D^*N$~\cite{GarciaRecio:2008dp}. Heavy
pseudoscalar and vector mesons should be treated on equal footing,
since they are degenerated in the heavy quark limit, and  are
connected by a spin-rotation of the heavy quark that leaves unaltered
the QCD Hamiltonian in that limit. This is to say the $D$ and $D^*$
mesons form a HQSS-doublet. 

The first molecular description of the
$\Lambda_c(2595)$ and $\Lambda_c(2625)$ resonances, using interactions fully consistent with HQSS, was derived in Refs.~\cite{GarciaRecio:2008dp, Romanets:2012hm}. In these works a consistent ${\rm SU(6)}_{\rm lsf} \times {\rm SU(2)}_{\rm HQSS}$
extension of the WT $\pi N$ Lagrangian --where ``lsf'' stands for
light-spin-flavor symmetry--, is implemented, although the adopted renormalization scheme (RS) \cite{Hofmann:2005sw, Hofmann:2006qx}  might not respect HQSS (see the discussion below).   Within such scheme, two
states are dynamically generated in the region of 2595 MeV. The first one, identified with the
$\Lambda_c(2595)$ resonance, is narrow and it strongly couples to 
$DN$ and especially to $D^*N$, with a small coupling to the open
$\pi\Sigma_c$ channel.  The second state is quite broad since it has a
sizable coupling to this latter channel. On the other hand, a
$J^P=3/2^-$ state is generated mainly by the $(D^*N-\pi\Sigma_c^*)$
coupled-channel dynamics. It would be the charm counterpart of the
$\Lambda(1520)$, and could be identified with the $\Lambda_c(2625)$ resonance. The same ${\rm SU(6)}_{\rm lsf} \times {\rm
  SU(2)}_{\rm HQSS}$ scheme also dynamically generates the
$\Lambda_b(5912)$ and $\Lambda_b(5920)$ narrow resonances, discovered
by LHCb in 2012 \cite{Aaij:2012da}, which turn out to be HQSS
partners, naturally explaining in this way their approximate mass
degeneracy~\cite{GarciaRecio:2012db}. The extension of the model to
the hidden charm  sector was carried out in
\cite{Garcia-Recio:2013gaa}, and more recently, it was
shown~\cite{Nieves:2017jjx} that some (probably at
least three) of the narrow $\Omega_c^*$ states recently observed by
LHCb~\cite{Aaij:2017nav} in the $\Xi_c^+K^-$ spectrum in $pp$
collisions  can be also dynamically
generated within the same scheme.

Several $\Lambda_c^*$ poles were also obtained in the approach followed in
Ref.~\cite{Liang:2014kra}. There, the interaction of $DN$ and $D^*N$ states, together with their coupled
channels are considered by using an extension to four flavours  of the SU(3) local hidden gauge formalism  from the light meson sector~\cite{Bando:1984ej,Bando:1987br,
  Meissner:1987ge}. The scheme also respects LO HQSS constraints~\cite{Xiao:2013yca} and, as in Refs.~\cite{GarciaRecio:2008dp,
  Romanets:2012hm}, a two-pole structure for the $\Lambda_c(2595)$ was also found, with the $D^*N$ channel playing a crucial role in
its dynamics. This is a notable
difference to the situation in the strange sector, where the analog
$\bar K^*N$ channel is not even considered in most of the studies of
the $\Lambda(1405)$, because of the large $\bar K^*-\bar K$ mass
splitting. (See also the discussion carried out in Ref.~\cite{Khemchandani:2011mf}.)

The beauty $\Lambda_b(5912)$ and $\Lambda_b(5920)$ states were also studied in the extended local hidden gauge (ELHG) approach in Ref.~\cite{Liang:2014eba}, while the the predictions of this scheme referred to the LHCb $\Omega_c^*$ states can be found in \cite{Liang:2017ejq}. These latter states were also addressed in Ref.~\cite{Montana:2017kjw} using a model constructed out of the 
SU(4)-flavor $t$-channel exchange of vector mesons. There, the original model of Ref.~\cite{Hofmann:2005sw} is revisited, 
and after taking an appropriate regularization scheme with physically sound parameters, two of the LHCb $\Omega_c^*$ resonances 
could be accommodated.  

\subsection{HQSS structure of the  $\Lambda_c(2595)$ and $\Lambda_c (2625)$ hadron-molecules}
\label{sec:hadmol}
To make more transparent the inner HQSS structure of the $\Lambda_{c\, (n)}^{\rm MOL}(2595)$, 
$\Lambda_{c\, (b)}^{\rm MOL}(2595)$ and $\Lambda_c^{\rm MOL}(2625)$
states found in molecular (MOL) scenarios [$(n)$ and $(b)$ refer to the
  narrow and broad resonances that form the two-pole structure of the
  $\Lambda_c(2595)$ in these schemes], we perform a change of
basis. We pass from $S-$wave states where the meson and baryon spins
are defined, to other ones, where the total angular momentum of the {\it ldof} is well determined. In both set  of states, the total angular momentum of the meson-baryon pair is defined. The two basis are related by a Racah rotation~\cite{Xiao:2013yca}, which is straightforward to obtain in 
the present case, where the discussion is restricted to $S-$wave meson-baryon pairs.  Thus for instance, we find (the rotation is independent of the isospin of the meson-baryon pair)
%
\begin{eqnarray}
|S_Q^P=1/2^+\, j_q^P=1^-; J^P = 3/2^- \rangle_1 & = &  |\pi \Sigma_c^*; J^P = 3/2^- \rangle\\
|S_Q^P=1/2^+\, j_q^P=1^-; J^P = 3/2^- \rangle_2 & = &  |D^* N; J^P = 3/2^- \rangle \\
|S_Q^P=1/2^+\, j_q^P=1^-; J^P = 1/2^- \rangle_1 & = &  |\pi \Sigma_c; J^P = 1/2^- \rangle \\
|S_Q^P=1/2^+\, j_q^P=1^-; J^P = 1/2^- \rangle_2 & = & \frac{\sqrt{3}}{2} |D N; J^P = 1/2^- \rangle + 
\frac12 |D^* N; J^P = 1/2^- \rangle \\
|S_Q^P=1/2^+\, j_q^P=0^-; J^P = 1/2^- \rangle_2 & = & -\frac12 |D N; J^P = 1/2^-\rangle + 
\frac{\sqrt{3}}{2} |D^* N; J^P = 1/2^- \rangle \label{eq:relation}
\end{eqnarray}
%
where we have used that the total angular momentum and parity of the {\it ldof} in the $\Sigma_c^{(*)}$ and $D^{(*)}$ ground states are $j_q^P=1^+$ and $1/2^-$, respectively. Besides, the sub-indices 1 and 2 on the states in the left-hand side of the equations distinguish if the meson is a Goldstone or a charmed heavy-light boson.
In this context, the approximate HQSS of QCD leads  to meson-baryon interactions $V$ satisfying (kinetic terms respect HQSS)
%
\begin{equation}
_\alpha\langle S_Q^P=1/2^+\, j_q^P; J^P | V | S_Q^P=1/2^+\, \hat j_q^{P'}; \hat J^{P'} \rangle_\beta = \delta_{j_q^P \hat j_q^{P'}}
\delta_{J^P \hat J^{P'}}\langle \alpha || V || \beta \rangle_{j_q^P}, \quad \alpha,\beta=1,2 \label{eq:meHQSS}
\end{equation}
%
where ${\cal O}(\Lambda_{\rm QCD}/m_Q)$ corrections have been neglected. The reduced matrix elements 
depend only on the configuration of the {\it ldof}, because QCD dynamics is invariant under  
spin rotations of the heavy quark in the infinite mass limit. Note that quantum numbers like isospin or strangeness \ldots, are conserved by QCD, and that for simplicity,  such trivial dependencies are not explicitly shown in Eq.~(\ref{eq:meHQSS}), though the $\langle \alpha || V || \beta \rangle_{j_q^P}$ elements obviously depend on these additional properties needed to  define the {\it ldof}.  Finally, just mention that, in principle, the orthogonal $|{j_q^P=1^-; J^P} \rangle_1$ and $|{j_q^P=1^-; J^P} \rangle_2$ states can be connected by an interaction respecting HQSS. For instance, in the context of models based on the exchange of vector mesons, these contributions  
necessarily involve a $D^*$,  instead of a $\rho-$meson, that will induce  
the transfer of charm between the baryon-baryon and meson-meson vertices.  

\subsubsection{${\rm SU(6)}_{\rm lsf} \times {\rm SU(2)}_{\rm HQSS}$}
To illustrate the discussion on the HQSS structure of the  $\Lambda_c(2595)$ and $\Lambda_c (2625)$ within molecular descriptions, we will focus on the  model derived in Refs.~\cite{GarciaRecio:2008dp, Romanets:2012hm}. There, the isoscalar interaction, $\widehat V$, used as kernel of the BSE in the $J^P=1/2^-$ and $J^P=3/2^-$ sectors respects HQSS (Eq.~(\ref{eq:meHQSS}))  and it leads to\footnote{Note that the order baryon-meson, instead of meson-baryon, is used in Refs.~\cite{GarciaRecio:2008dp, Romanets:2012hm}. This induces a minus sign for off diagonal elements involving the $D^*N$ pair in the $J=1/2$ sector. In addition,  there exists a minus sign of difference  between the conventions of  \cite{GarciaRecio:2008dp, Romanets:2012hm} and those adopted here for the $\Sigma_c^*$. \label{foot:sign}}
%
\begin{equation}
\langle 1  || \widehat V|| 1 \rangle_{1^-} = -4f(s), \quad \langle 2  || \widehat V || 2 \rangle_{1^-} = 0, \quad \langle 1  || \widehat V || 2 \rangle_{1^-} = \sqrt{2}f(s), \quad \langle 2 ||\widehat V || 2 \rangle_{0^-} = -12f(s), \label{eq:we-me}
\end{equation}
%
when the coupled-channels space is truncated to that generated by the $\pi\Sigma^{(*)}_c$ and $D^{(*)}N$ pairs. Besides, $f(s)$ is a function of the meson-baryon Mandelstam variable $s$. Note that $\langle 1  || \widehat V|| 1 \rangle_{1^-}$ is determined by the isoscalar $\pi\Sigma_c^{(*)}\to\pi \Sigma_c^{(*)}$ transition, which  is approximated in ~\cite{GarciaRecio:2008dp, Romanets:2012hm} by the 
LO WT chiral interaction. This fixes $f(s)$ to 
\begin{equation}
f(s)= \frac{\sqrt{s}-M}{2f^2_\pi}\, \frac{E+M}{2M}
\end{equation}
using the normalizations of these works. In the above equation,  $M(E)$ is the common mass [center-of-mass
energy] of the $\Sigma_c^{(*)}$ baryons and $f_\pi \sim 92$ MeV is the pion decay constant\footnote{In the approach of  
Refs.~\cite{GarciaRecio:2008dp, Romanets:2012hm} sizable flavor symmetry breaking terms are included. Actually, the symmetry-pattern exhibited by the reduced matrix elements  in Eq.~(\ref{eq:we-me}) is modified, by computing the function $f(s)$ using  physical
hadron masses and decay constants (see for instance, Eq.~(7) of Ref.~\cite{Romanets:2012hm}). This induces  mostly SU(4)-flavor 
breaking corrections, since the charmed-hadrons masses and decay constants follow in good approximation the HQSS-predictions, which  do not significantly alter the discussion that follows.}. Coming back to Eq.~(\ref{eq:we-me}), we see a large attraction  for the $j_q^P=0^-$ {\it ldof} configuration, which is constructed out of the $DN$ and $D^*N$ pairs, since the {\it ldof} in the $S-$wave $\pi\Sigma_c$ channel can be only  $j_q^P=1^-$. Indeed, the $j_q^P=0^-$  eigenvector of 
the matrix $\widehat V $ is
%
\begin{equation}
v_0^{\rm atr}\equiv|S_Q^P=1/2^+\, j_q^P=0^-; J^P = 1/2^- \rangle_2 \, .
\end{equation}
%
On the other hand, diagonalizing $\widehat V $  in the $j_q^P=1^-$ {\it ldof} subspace, we find additional attractive and slightly repulsive eigenvalues $\lambda_1^{\rm atr}=-2-\sqrt{6}\sim -4.45 $ and  $\lambda_1^{\rm rep}=-2+\sqrt{6}\sim 0.45$, respectively, to be compared to  $\lambda_0=-12$ obtained in the $j_q^P=0^-$ sector. The corresponding eigenvectors are $v_1^{\rm atr}\sim (1,\sqrt{2}-\sqrt{3})$ and $v_1^{\rm rep}\sim(\sqrt{3}-\sqrt{2}, 1)$ in the $ |j_q^P=1^- \rangle_\alpha,\, \alpha=1, 2$ basis.  Taking normalized vectors, we find for $J^P=1/2^-$
\begin{equation}
 ||v_1^{\rm atr}||^2_{1/2^-} =  \underbrace{0.91}_{\pi\Sigma_c}+\underbrace{0.07}_{DN}+\underbrace{0.02}_{D^*N}\,, \quad
  ||v_1^{\rm rep}||^2 _{1/2^-}=  \underbrace{0.09}_{\pi\Sigma_c}+\underbrace{0.68}_{DN}+\underbrace{0.23}_{D^*N}\, ,\quad
  ||v_0^{\rm atr}||^2_{1/2^-} =  \underbrace{0.25}_{DN}+\underbrace{0.75}_{D^*N}\, , \label{eq:norms1}
\end{equation}
while for $J^P=3/2^-$, we have
\begin{equation}
 ||v_1^{\rm atr}||^2_{3/2^-} = \underbrace{0.91}_{\pi\Sigma_c^*}+\underbrace{0.09}_{D^*N}\, , \quad
  ||v_1^{\rm rep}||^2 _{3/2^-}=  \underbrace{0.09}_{\pi\Sigma_c^*}+\underbrace{0.91}_{D^*N} \label{eq:norms2}
\end{equation}
In light of these results, we could easily explain some features of the results found in Refs.~\cite{GarciaRecio:2008dp, Romanets:2012hm} for the lowest-lying 
odd-parity $\Lambda_c^*$ states. There, a narrow $J^P=1/2^-$ $\Lambda_{c\, (n)}^{\rm MOL}(2595)$ resonance ($\Gamma \sim 1$ MeV) is reported, mostly generated from the extended WT $DN-D^*N$ coupled-channels dynamics. The modulus square of the couplings of this resonance to $DN$ and $D^*N$ are approximately in the ratio 1 to 2.4, which does not differ much from the 1 to 3,  that one would expect from the decomposition of $||v_0^{\rm atr}||^2_{1/2^-}$ in Eq.~(\ref{eq:norms1}). Besides, this state has a small coupling to the $\pi\Sigma_c$ channel, which further supports a largely dominant $0^-$ {\it ldof} attractive configuration in its structure. Moreover, the detailed analysis carried out in  \cite{Romanets:2012hm} reveals that this narrow resonance stems from a ${\bf 21}$ $\rm SU(6)_{ lsf}$ irreducible representation (irrep), where the light quarks --three quarks and anti-quark--  behave (do not behave)  as an isoscalar 
spin-singlet (triplet) diquark--symmetric spin-flavor state--. 

The RS adopted in Ref.~\cite{GarciaRecio:2008dp, Romanets:2012hm}, proposed in \cite{Hofmann:2005sw, Hofmann:2006qx}, plays an important role in enhancing the influence of the $D^*N$ channel in the dynamics of the narrow $\Lambda_{c\, (n)}^{\rm MOL}(2595)$ state. Furthermore, this RS also produces a reduction in the mass of the resonance of around 200 MeV, which thus appears in the region of 2.6 GeV, instead of in the vicinity of the $DN$ threshold. 
The RS establishes that all loop functions are set to zero at a common point [$\mu= \sqrt{m_{\rm th}^2+M_{\rm th}^2}$, where $(m_{\rm th}+M_{\rm th})$ is the mass of the lightest hadronic channel], regardless of the total angular moment $J$ of the sector. However, we should point out that  such RS might not be fully consistent with HQSS. 

In addition, there appears a second $J^P=1/2^-$ pole [$\Lambda_{c\, (b)}^{\rm MOL}(2595)$]  in the 2.6 GeV region~\cite{GarciaRecio:2008dp, Romanets:2012hm}. Although it is  placed relatively close to the $\pi\Sigma_c$ threshold, this resonance is broad ($\Gamma \sim 70-90$ MeV) thanks to its sizable coupling to this open channel, which  in this case is larger than  those to  $DN$ and $D^*N$. The study of Ref.~\cite{Romanets:2012hm} associates this isoscalar resonance to a ${\bf 15}$ $\rm SU(6)_{ lsf}$ irrep, where the {\it ldof}  effectively behave as an isoscalar spin-triplet diquark (antisymmetric spin-flavor configuration). Thus, it is quite reasonable to assign a dominant $j_q^P=1^-$ configuration to the {\it ldof} in this second pole. However, the ratios of $\pi\Sigma_c, DN$ and $D^*N$ couplings of this second resonance do not follow the pattern inferred from $||v_1^{\rm atr}||^2 _{1/2^-}$ in Eq.~(\ref{eq:norms1}) as precisely as in the case of the narrow state. Actually,  the couplings of this broad state to the $DN$ and $D^*N$ pairs, though smaller, turn out to be comparable (absolute value) in magnitude to the $\pi \Sigma_c$ one (1.6, 1.4 and  2.3, respectively~\cite{Romanets:2012hm}). This points to the possibility that this second pole might also have a sizable component of the $1^-$ repulsive configuration, for which we should expect  $DN$ and $D^*N$ couplings much larger than the $\pi \Sigma_c$ one (likely in proportion 9 to 1 for the squares of the absolute values, just opposite to what is expected from the $1^-$ attractive eigenvector in Eq.~(\ref{eq:norms1})). Indeed, the fact  
that the $\Lambda_{c\, (b)}^{\rm MOL}(2595)$ is located above the $\pi\Sigma_c$ threshold reinforces this picture, where there would be a significant mixing among the attractive and repulsive $1^-$ configurations, provoked  by the flavor breaking corrections incorporated in the model of 
Refs.~\cite{GarciaRecio:2008dp, Romanets:2012hm}. These symmetry breaking terms affect  the kernel $f(s)$ of the BSE, the meson-baryon loops and   the renormalization of the UV behaviour of the latter to render finite the unitarized amplitudes. The large difference between the actual $\pi \Sigma_c$   and $D^{(*)}N$ thresholds, which are supposed to be degenerate to obtain the results of Eq.~(\ref{eq:norms1}), should certainly play an important role. The mass breaking effects were less relevant for the narrow $\Lambda_{c\, (n)}^{\rm MOL}(2595)$ resonance, because in that case  i) the $\pi \Sigma_c$ channel had  little influence in the dynamics of the state, and ii) the dominant $DN$ and $D^*N$ thresholds turn out to be relatively close, thanks to HQSS. In addition,  other higher channels like $\eta\Lambda_c$, $K \Xi^{(\prime)}_c $, $D_s \Lambda$, $\rho \Sigma_c, \ldots$ which are considered in \cite{GarciaRecio:2008dp, Romanets:2012hm}, have not been included here 
in the simplified analysis that leads to  the results of Eq.~(\ref{eq:norms1}). Finally, one should neither discard a small $0^-$ {\it ldof} 
component in the $\Lambda_{c\, (b)}^{\rm MOL}(2595)$ wave-function that will also change the couplings of this broad state to the different channels.

Note that the total angular momentum and parity of the {\it ldof} are neither really conserved in the ${\rm SU(6)}_{\rm lsf} \times {\rm SU(2)}_{\rm HQSS}$ model, nor  in the real physical world  because the charm quark mass is finite. Hence, both the narrow and broad $\Lambda_{c\, (n,b)}^{\rm MOL}(2595)$ resonances reported in \cite{GarciaRecio:2008dp, Romanets:2012hm} will have an admixture of the $0^-$ and $1^-$ configurations\footnote{However, the previous discussion has allowed us to reasonably identify the dominant one in each case. The existence of a certain  mixing is out of doubt, thus for instance, the narrow state can decay into $\pi\Sigma_c$ through its  $1^-$ small component.} in their inner structure.  More importantly, the physical $\Lambda_c(2595)$ and the second resonance, if it exists, will also contain both type of {\it ldof} in their wave-function. As stressed in the Introduction, a non-negligible $0^-$ component in the  $\Lambda_c(2595)$ or a double-pole structure have not been considered in the theoretical analyses of the 
exclusive semileptonic $\Lambda_b$ decays into $\Lambda_c(2595)$ carried out in Refs.~\cite{Roberts:1992xm, Leibovich:1997az,Boer:2018vpx}. One of the main objectives of this work is precisely the study of how these non-standard features affect the $\Lambda_b\to \Lambda_c^*$ transitions. 
     
Finally, the lowest-lying $J^P=3/2^-$ isoscalar resonance found in
Refs.~\cite{GarciaRecio:2008dp, Romanets:2012hm} is clearly the HQSS
partner of the broad  $J^P=1/2^-$ $\Lambda_{c\, (b)}^{\rm MOL}(2595)$
state, with quantum number  $j_q^P=1^-$ for the {\it ldof}. It is located above the $\pi \Sigma^*_c$ threshold, with a width of around 40-50 MeV, and placed in the ${\bf 15}$ $\rm SU(6)_{ lsf}$ irrep~\cite{Romanets:2012hm}, as the broad  $\Lambda_{c\, (b)}^{\rm MOL}(2595)$ resonance. Moreover, 
the complex coupling of this $J^P=3/2^-$ pole to the $\pi\Sigma_c^*$ channel is essentially identical to that of the $\Lambda_{c\, (b)}^{\rm MOL}(2595)$ to $\pi\Sigma_c$. In turn, the square of the absolute value of its coupling to $D^*N$ compares reasonably well with the sum of the squares of the couplings of the $\Lambda_{c\, (b)}^{\rm MOL}(2595)$  to $DN$ and $D^*N$, as one would expect from Eqs.~(\ref{eq:norms1}) and (\ref{eq:norms2}).   This $J^P=3/2^-$ isoscalar resonance is identified  with the $D-$wave $\Lambda_c(2625)$ in Refs.~\cite{GarciaRecio:2008dp, Romanets:2012hm}. In these works, it is argued that 
a small change in the renormalization subtraction constant  could easily move the resonance down by 40 MeV to the nominal position of the physical state, and that in addition, this change of the mass  would considerably reduce the width, since its position would get much closer to the
threshold of the only open channel $\pi \Sigma_c^*$.

Thus, within the ${\rm SU(6)}_{\rm lsf} \times {\rm SU(2)}_{\rm HQSS}$ model, the $\Lambda_c(2625)$ turns out to be the HQSS partner of the second broad 
$\Lambda_{c\, (b)}^{\rm MOL}(2595)$ pole instead of the narrow  $\Lambda_{c\, (n)}^{\rm MOL}(2595)$ resonance, as commonly assumed in the theoretical analyses of the 
exclusive semileptonic $\Lambda_b$ decays into $\Lambda_c(2595)$. This picture clearly contradicts the predictions of the CQMs where first, there is no a second 2595 pole, and second, the $\Lambda_c(2625)$ and the narrow $\Lambda_c(2595)$ are HQSS siblings, produced by a $\lambda-$mode excitation of the ground $1/2^+$ $\Lambda_c$ baryon.

\subsubsection{Extended local hidden gauge (ELHG)}

Within the model of Ref.~\cite{Liang:2014kra}, the dynamics of the lowest-lying odd-parity $\Lambda_c^*$ is mostly governed by the $DN$, $D^*N$ and $\pi\Sigma_c$ interactions ($V^{\rm HG}$). They are constructed  using an SU(4) extension of the local hidden gauge formalism derived for the light meson sector~\cite{Bando:1984ej,Bando:1987br,  Meissner:1987ge}, that in a first stage respects HQSS. It gives rise to reduced matrix elements 
\begin{equation}
\langle 1  || V^{\rm HG}|| 1 \rangle_{1^-} = -4f(s), \quad \langle 2  ||  V^{\rm HG} || 2 \rangle_{1^-} = -3f(s), \quad \langle 1  ||  V^{\rm HG} || 2 \rangle_{1^-} = 0, \quad \langle 2 ||V^{\rm HG} || 2 \rangle_{0^-} = -3f(s), \label{eq:we-me-hg}
\end{equation}
in the isoscalar sector. The flavor symmetry of the WT function $f(s)$  is broken in the meson-baryon  space by the use of physical masses. At first, $D^*-$exchange driven interaction terms connecting   $|{j_q^P=1^-; J^P} \rangle_1$ and $|{j_q^P=1^-; J^P} \rangle_2$ states are neglected, 
as well as $DN\to D^*N$ coupled-channel interactions in the $J^P=1/2^-$ sector. 

In a second stage, some additional contributions driven by the $D^*D\pi$ coupling, that formally vanish in the infinitely heavy quark mass limit, are considered in the kernels (potentials) of the BSE. These new terms provide: 
\begin{itemize}

\item First, $DN\to \pi \Sigma_c$ transitions in the $J^P=1/2^-$ sector, which would give rise to $\quad \langle 1  || V^{\rm HG} || 2 \rangle_{1^-} = \sqrt{2}f(s)/4$. The factor $1/4$ roughly accounts for the ratio $(m_\rho/m_{D^*})^2$,  which one would expect to suppress the diagrams induced by the $t-$channel exchange of charmed vector mesons compared to  those mediated by members of the light $\rho-$octet~\cite{Mizutani:2006vq}. This assumes  
a universal KSFR vector-meson coupling. However, the effects due to $\quad \langle 1  || V^{\rm HG} || 2 \rangle_{1^-}\neq 0$  are, inconsistently with HQSS,  not considered in the $J^P=3/2^-$ sector, and thus $D^*N$ and $\pi\Sigma_c^*$ channels are not connected\footnote{We should also point out that 
the $D^*N\to \pi \Sigma_c$ transition in the $J^P=1/2^-$ sector is also set to zero in ~\cite{Liang:2014kra}. This is also inconsistent with HQSS, since this symmetry relates this off diagonal term of the interaction  with the $ DN \to \pi \Sigma_c $ one (a factor $ 1/\sqrt3$).} in the formalism of Ref.~\cite{Liang:2014kra}. Actually, the isoscalar $\pi\Sigma_c^*$ pair is separately treated as a single channel. We will come back to this point below. 

\item Second, $D^{(*)}N\to  D N$ transitions in the $J^P=1/2^-$ sector obtained from box diagrams, which also generate  contributions to the $DN\to  DN$ and $D^*N\to  D^*N$ diagonal interaction-terms. In the $J^P=3/2^-$ sector, modifications of the  $D^*N\to  D^*N$ potential induced by box-diagrams constructed out, in this case, of the anomalous $D^*D^*\pi$ coupling are also taken into account in ~\cite{Liang:2014kra}.

\end{itemize} 
In addition,  other higher channels like  $\eta\Lambda_c$, $\rho \Sigma_c, \ldots$ are considered in \cite{Liang:2014kra}, though they have a little influence in the lowest-lying $\Lambda_c^*$ states. After fine tuning  some UV cutoffs to reproduce the masses of the experimental narrow  $\Lambda_c(2595)$ and $\Lambda_c(2625)$, the authors of  Ref.~\cite{Liang:2014kra} found that the latter resonance is essentially a $D^*N$ state, while the former one couples strongly both to $DN$ and $D^*N$ and has  a quite small coupling to $\pi\Sigma_c$. In addition, a state at $2611$ MeV and a width of around 100 MeV, which couples mostly to $\pi\Sigma_c$ is also dynamically generated, confirming the double pole structure predicted in   the ${\rm SU(6)}_{\rm lsf} \times {\rm SU(2)}_{\rm HQSS}$ model 
of Refs.~\cite{GarciaRecio:2008dp, Romanets:2012hm}. Note also that the narrow  $\Lambda_c(2595)$ state found in \cite{Liang:2014kra} has similar $DN$ and $D^*N$ couplings,  from where one can conclude that it should have an important $0^-$ {\it ldof} configuration.

On the other hand, in the $J^P=3/2^-$ sector the isoscalar $\pi\Sigma_c^*$ is treated as a single channel in \cite{Liang:2014kra}. It gives rise to a further broad state ($\Gamma\sim$ 100 MeV) in the region of 2675 MeV, which is 
not related to the $\Lambda_c(2625)$ in that reference.

Finally, we should mention that the box-diagrams interaction terms evaluated in this ELHG model  break  HQSS 
at the charm scale, and it becomes difficult to identify any HQSS resonance doublet among the results reported in \cite{Liang:2014kra}. 

\subsubsection{SU(4) flavor $t$-channel exchange of vector mesons}
\label{sec:su4}
As already mentioned in this kind of models ~\cite{Hofmann:2005sw, Mizutani:2006vq},  the BSE potentials are calculated from the zero-range limit of $t-$channel exchange of vector mesons between pseudoscalar mesons and baryons. Chiral symmetry is preserved in the light
meson sector, while the interaction is still of the WT type. Thus, the $J=1/2$ lowest-lying odd-parity $\Lambda_c^*$ resonances are mostly generated from $DN, \pi \Sigma_c$ coupled-channels dynamics. SU(4) flavor symmetry is used to determine the 
$DN\to DN$ and $DN\to \pi\Sigma_c$ interactions, which could be also derived assuming that the KSFR coupling relation 
holds also when charm hadrons are involved. The flavor symmetry is broken by the physical hadron masses, and in particular the large 
mass of the $D^*$ suppresses the off diagonal matrix element $DN\to \pi\Sigma_c$, as compared to the diagonal ones that are driven by $\rho-$meson exchange (see also the discussion in the previous subsection about the factor 1/4 included in the ELHG approach of Ref.~\cite{Liang:2014kra}). These approaches do not include the $D^*N\to D^*N, DN, \pi\Sigma_c$ transitions, and therefore are not consistent with HQSS. Nevertheless,  a  $J^P = 1/2^-$ narrow resonance 
close to the $\pi\Sigma_c$ threshold, which can be readily identified with
the $\Lambda_c(2595)$, is generated. It couples strongly to $DN$, and its nature is
therefore very different from those obtained in the ${\rm SU(6)}_{\rm lsf} \times {\rm SU(2)}_{\rm HQSS}$ and in the ELHG models, for 
which the $D^*N$ channel plays a crucial role. The reason why these SU(4) models can generate
the  $\Lambda_c(2595)$ is that the lack of the $D^*N$  in the $J^P=1/2^-$ sector is  compensated by the
enhanced strength in the $DN$ channel. For instance,  the $DN$  coupling in the approaches of Refs.~\cite{Hofmann:2005sw, Mizutani:2006vq} turned out to be of the same magnitude as that of the narrow  $\Lambda_{c\, (n)}^{\rm MOL}(2595)$ to $D^*N$ in the ${\rm SU(6)}_{\rm lsf} \times {\rm SU(2)}_{\rm HQSS}$  model of 
Refs.~\cite{GarciaRecio:2008dp, Romanets:2012hm}. On the other hand, the $\pi \Sigma_c$ coupling, though still small, was found twice larger  in  Refs.~\cite{Hofmann:2005sw, Mizutani:2006vq}.  By construction, the resonance described in  \cite{Hofmann:2005sw, Mizutani:2006vq} will mix $j_q^P=0^- $ and $1^-$ {\it ldof} configurations. The gross features of this dynamically generated state are similar to those of the the resonance reported in Ref.~\cite{Tolos:2004yg}, where the similarity between the $DN$ and $\bar K N$ systems, once the strange quark in the later is replaced by a charm quark, was exploited.

In addition, the models based on the $t$-channel exchange of vector mesons, when the unitarized amplitudes are renormalized as suggested in \cite{Hofmann:2005sw, Hofmann:2006qx}, produce also a second $J^P=1/2^-$ broad resonance ($\Gamma \sim 100$ MeV) above 2600 MeV, with $\pi\Sigma_c$ (largest) and $DN$ couplings similar to those found in the ${\rm SU(6)}_{\rm lsf} \times {\rm SU(2)}_{\rm HQSS}$ and in the ELHG approaches (see Table XIV of Ref.~\cite{GarciaRecio:2008dp} and the related discussion for an update of the results of the model used in \cite{Mizutani:2006vq}). Therefore, this type of molecular models might also predict a double pole structure for the $\Lambda_c(2595)$, in analogy with what happens  in the unitary chiral descriptions of the  $\Lambda(1405)$. We should, however, note that this second broad state is not generated when a RS based on 
an UV hard-cutoff is used~\cite{Mizutani:2006vq,JimenezTejero:2009vq}.

In the isoscalar $J^P=3/2^-$ sector, the chiral $\pi \Sigma_c^*$ WT interaction, driven by $\rho-$exchange, leads to a resonance with some resemblances to  that reported in Refs.~\cite{GarciaRecio:2008dp, Romanets:2012hm}, and that it is identified in \cite{Hofmann:2006qx} with the $\Lambda_c(2625)$, despite being located above 2660 MeV and having a width of the order of 50 MeV. Actually, this pole corresponds to that found in the single channel $\pi\Sigma_c^*$ analysis of 
Ref.~\cite{Liang:2014kra}, where it was, however,  not associated to the physical $\Lambda_c(2625)$ state.

\subsubsection{Chiral isoscalar $\pi\Sigma^{(*)}_c$ molecules}
\label{sec:chiral}

The chiral interactions between the ground-state singly charmed baryons and the Goldstone bosons lead to 
scenarios~\cite{Lutz:2003jw, Hofmann:2006qx, Lu:2014ina} where $\pi\Sigma_c$ and $\pi\Sigma^*_c$ isoscalar molecules naturally 
emerge in the $J^P=1/2^-$ and $J^P=3/2^-$ sectors, respectively. These states will form a $1^-$ HQSS doublet, whose masses and widths 
depend on the details of the used RS. The works of Refs.~\cite{Lutz:2003jw, Hofmann:2006qx} found $J^P=1/2^-, 3/2^-$  
resonances of around 50 MeV of width and masses in the 2660 MeV region using a  RS, inspired in the success of  Refs.~\cite{Lutz:2001yb, GarciaRecio:2003ks,  Kolomeitsev:2003kt} to describe the  chiral SU(3) meson-baryon $J^P=1/2^-$ and $J^P=3/2^-$ sectors, later also  employed in  the ${\rm SU(6)}_{\rm lsf} \times {\rm SU(2)}_{\rm HQSS}$ model of Refs.~\cite{GarciaRecio:2008dp, Romanets:2012hm}\footnote{As we discussed above, the consideration of the $DN$ and $D^*N$ channels in \cite{GarciaRecio:2008dp, Romanets:2012hm} strongly  modifies the $J^P=1/2^-$ sector, leading to a quasi-bound $D^*N$ state.}. The $\pi\Sigma_c^*$ pole found in the ELHG scheme followed in \cite{Liang:2014kra} clearly matches the results of Ref.~\cite{Hofmann:2006qx}, though  it was not identified with the $\Lambda_c(2625)$ in the work of Ref.~\cite{Liang:2014kra}. 

In sharp contrast, subtraction constants or UV cutoffs  were fine-tuned in Ref.~\cite{Lu:2014ina} in such a way  that the $\Lambda_c(2595)$ and  $\Lambda_c(2625)$ experimental masses were reproduced, leading to weakly $\pi\Sigma_c$ and $\pi\Sigma_c^*$ bound states. Thus, the needed UV cutoffs turned out to be slightly higher than expected, 1.35 and 2.13 GeV, respectively. This could indicate some degrees of freedom that are not considered in the approach, such that CQM states or $D^{(*)}N$ components, and that could play a certain role, being their effects effectively accounted for the fitted real parts of the unitarity loops~\cite{Guo:2016nhb, Albaladejo:2016eps}.

\subsection{Weinberg compositeness condition}

In recent years, the compositeness
condition, first proposed by Weinberg to explain the deuteron as a
neutron-proton bound state~\cite{Weinberg:1962hj,Weinberg:1965zz}, has
been advocated as a model independent way to determine the relevance
of hadron-hadron components in a molecular state. With renewed
interests in hadron spectroscopy, this method has been extended to
more deeply bound states, resonances, and higher partial
waves~\cite{Baru:2003qq,Cleven:2011gp,Gamermann:2009uq,YamagataSekihara:2010pj,
  Aceti:2012dd,Aceti:2014ala,Aceti:2014wka,Hyodo:2011qc,Hyodo:2013nka,Sekihara:2014kya,Nagahiro:2014mba,Garcia-Recio:2015jsa}. However, 
  we should mention that the compositeness analysis proposed by
Weinberg~\cite{Weinberg:1962hj,Weinberg:1965zz} is only valid
for bound states. For resonances, it involves complex numbers and,
therefore, a strict probabilistic interpretation is lost as pointed out in Ref.~\cite{Aceti:2014ala}.

For the particular case of the $\Lambda_c(2595)$, the situation is a
bit unclear. For instance, it was shown in Ref.~\cite{Hyodo:2013iga}
that the $\Lambda_c(2595)$ is not predominantly a $\pi\Sigma_c$
molecular state using the effective range expansion. A similar
conclusion was reached in Ref.~\cite{Guo:2016wpy}, using a generalized
effective range expansion including Castillejo-Dalitz-Dyson pole
contributions. In this latter work, the effects of isospin breaking
corrections are also taken into account and the extended compositeness
condition for resonances developed in Ref.~\cite{Guo:2015daa} was 
applied to calculate the component coefficients. Furthermore,
although in the unitary approaches,
the $\Lambda_c(2595)$ is found to be of molecular nature~\cite{Lutz:2003jw,Tolos:2004yg, Mizutani:2006vq, Hofmann:2005sw,GarciaRecio:2008dp,Romanets:2012hm,Liang:2014kra,Lu:2014ina}, there is no
general agreement on its dominant meson-baryon components yet. 

In general, one can conclude that the compositeness of the $\Lambda_c(2595)$ depends on the number of considered coupled channels, 
and on the particular regularization scheme adopted in the unitary
approaches and, therefore, would be model dependent~\cite{Lu:2016gev}.

\section{Semileptonic $\Lambda_b \to \Lambda_c^*\ell\bar{\nu}_\ell$ decays}
\label{sec:SL-decay}
\begin{figure}[tbh]
\includegraphics[clip,trim=3cm 17cm 4cm 3cm, scale=0.6]{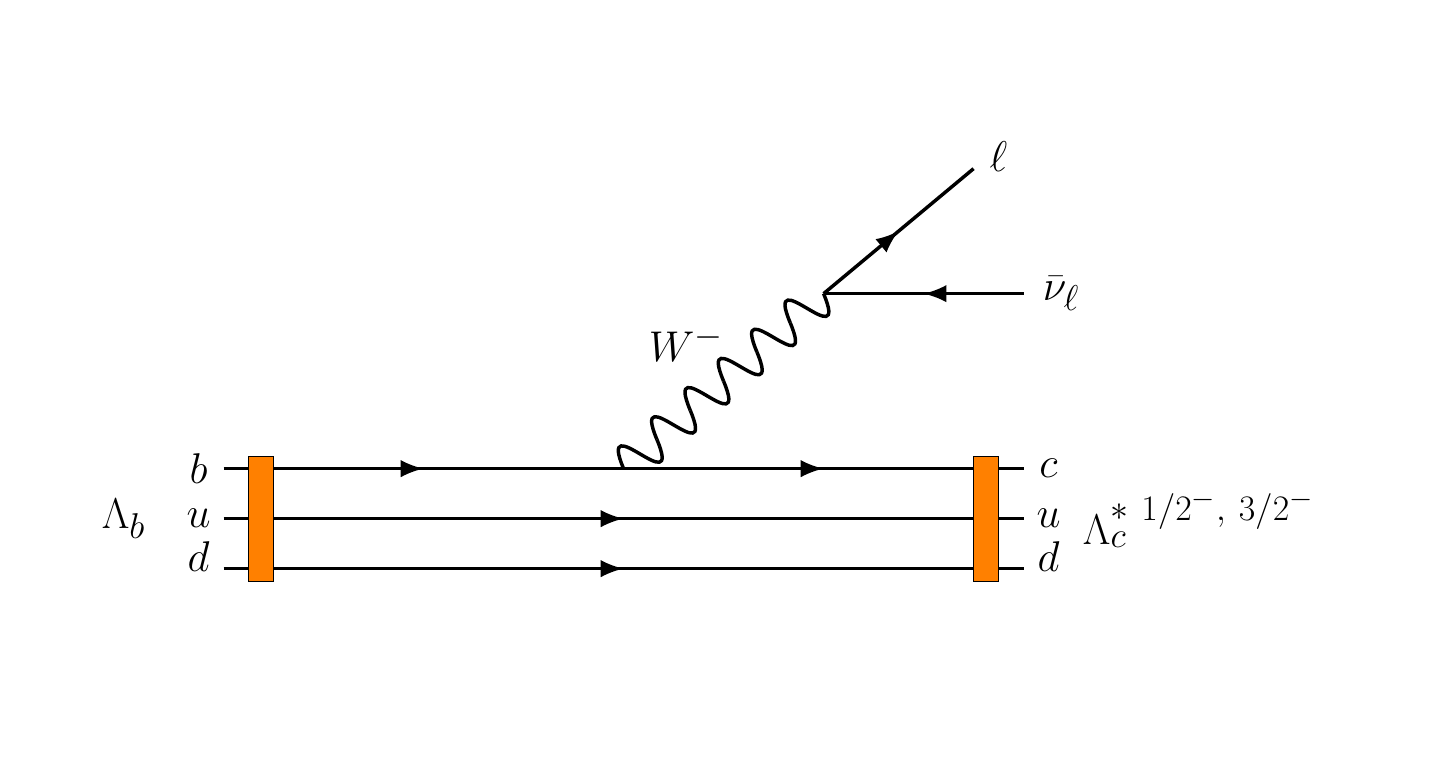} 
\includegraphics[clip,trim=3cm 17cm 4cm 3cm, scale=0.6]{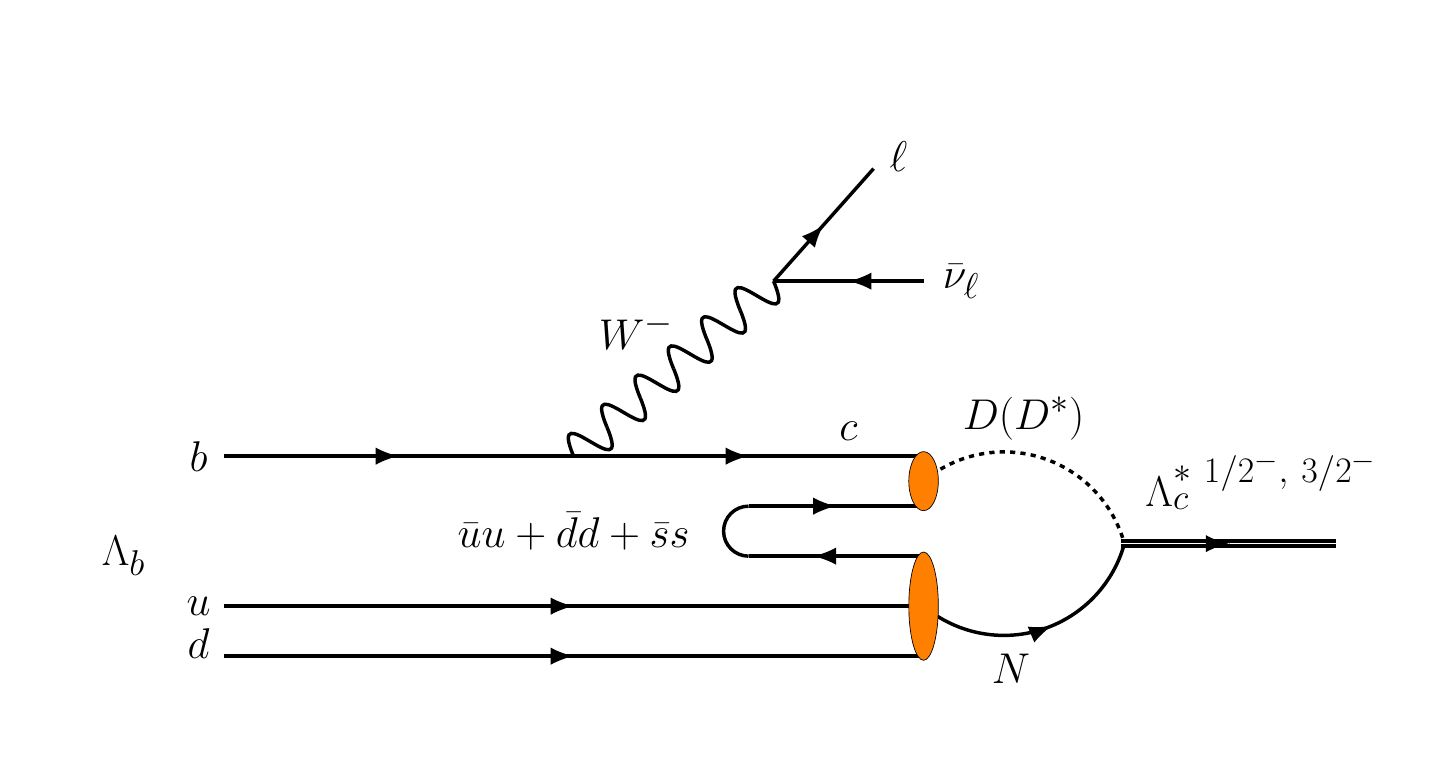} 
\caption{Left: Diagrammatic representation of the $\Lambda_b \to \Lambda_c^*\ell\bar{\nu}_\ell$ decay. Right: Hadronization creating $q\bar q$ pairs, together with the pictorial representation of the mechanism to produce a $\Lambda^*_c$ resonance,  through an intermediate propagation
of $DN$ and $D^*N$ pairs. }
 \label{fig:SL}
\end{figure}
%
The differential decay width for the semileptonic $b\to c$ transition shown in Fig.~\ref{fig:SL} is given by
\begin{equation}
\frac{d\Gamma}{d\omega}= \Gamma_0 
 \frac{96\,M_{\Lambda_c^*}^3}{\pi\,M_{\Lambda_b}^5} \sqrt{\omega^2-1}\, {\cal
L}^{\alpha\beta}(q) {\cal H}_{\alpha\beta}(P,P')\,, \qquad \Gamma_0 =  |V_{cb}|^2 \frac{G_F^{\,2}M_{\Lambda_b}^5}{192\pi^3}
\end{equation}
where $|V_{cb}|$ is the modulus of the Cabibbo--Kobayashi--Maskawa (CKM) matrix element for the $b\to c$ transition, $G_F= 1.16638 \times 10^{-11}$\,MeV$^{-2}$ is the Fermi decay constant, $P^\mu,M_{\Lambda_b}$ ($P^{\prime \mu},M_{\Lambda_c^*}$) are the four-momentum and mass of
the initial (final) baryon, $q^\mu=P^\mu-P^{\prime \mu}$ and $\omega$ is the product of the
baryons four-velocities $[P^{(\prime)\mu}/M_{\Lambda_{b,c}^{(*)}}]$, 
$\omega=v\cdot v^\prime =\frac{M_{\Lambda_b}^2+M_{\Lambda_c^*}^2-q^2}{2M_{\Lambda_b}M_{\Lambda_c^*}}$. In the
decay, $\omega$ ranges from $\omega=1$, corresponding to zero recoil of the
final baryon, to a maximum value given, neglecting the antineutrino mass, by $\omega=\omega_{\rm max}= \frac{M_{\Lambda_b}^2 + M_{\Lambda_c^*}^2-m_\ell^2}{2M_{\Lambda_b}M_{\Lambda_c^*}}$, where $m_\ell$ is the final charged lepton mass. In addition, 
${\cal L}^{\alpha\beta}(q)$ is the leptonic tensor after integrating
in the lepton momenta
\begin{eqnarray}
{\cal L}^{\alpha\beta}(q)&=& \int\frac{d^3k}{2|\vec{k}\,|}\frac{d^3k'}{2\sqrt{m^2_{\ell}+\vec{k}'^2}}\left( k'^\alpha k^\beta +k'^\beta k^\alpha
- g^{\alpha\beta} k\cdot k^\prime + {\rm i}
\epsilon^{\alpha\beta\rho\sigma}k'_{\rho}k_\sigma \right)  \delta^4(q-k-k')\nonumber\\
&=&-\frac{\pi }{6 q^2}(q^2-m^2_\ell)\left\{\left(q^2-\frac{m_\ell^2}{2}-\frac{m_\ell^4}{2q^2}\right)\,g^{\alpha\beta}-
\left(q^2+m_\ell^2-2\frac{m_\ell^4}{q^2}\right)\,\frac{q^\alpha q^\beta}{q^2}\right\}
\end{eqnarray}
where $k$ and $k'$ are the four-momenta of the outgoing antineutrino and charged lepton [in our convention, we take $\epsilon_{0123}=+1$ 
and the metric $g^{\mu\mu}=(+,-,-,-)$]. Besides, ${\cal H}_{\alpha\beta}(P,P')$ is the
hadronic tensor given by
\begin{eqnarray}
{\cal H}^{\alpha\beta}(P,P') &=& \frac{1}{2} \sum_{r,r'}  
 \big\langle \Lambda_c^{*}, r'\
\vec{P}^{\,\prime}\big| J_{bc}^\alpha(0)\big| \Lambda_b, r\ \vec{P}   \big\rangle 
\ \big\langle \Lambda_c^{*}, r'\ 
\vec{P}^{\,\prime}\big|J_{bc}^\beta(0) \big|  \Lambda_b, r\ \vec{P} \big\rangle^*
\label{eq:wmunu}
\end{eqnarray}
with $\big|\Lambda_b, r\ \vec P\big\rangle\,
\left(\big|\Lambda_c^{*}, r'\ \vec{P}\,'\big\rangle\right)$ the initial (final)
baryon state with three-momentum $\vec P$ ($\vec{P}\,'$) and helicity $r$ ($r'$), and normalized such that 
\begin{equation}
\big\langle B, r'\
\vec{P}'\, |\,B, r \ \vec{P} \big\rangle = (2\pi)^3 \frac{E}{M}
\,\delta_{rr'}\, \delta^3 (\vec{P}-\vec{P}^{\,\prime}), \quad B= \Lambda_b, \Lambda_c^{*}
\end{equation}
with $E$ and $M$, the baryon energy for three-momentum $\vec P$ and its mass, respectively. Finally, $J_{bc}^\mu(0)$ is
the $b\to c $ charged weak current 
\begin{equation}
J_{bc}^\mu(0)=\bar\Psi_{c}(0)\gamma^\mu(1-\gamma_5)\Psi_b(0) 
\end{equation}
with $\Psi_{b,c}$, Dirac fields,  with dimensions of mass to the 3/2.
Hadronic matrix elements  can be parameterized in terms of form factors~\cite{Leibovich:1997az}.
For $\frac12^+ \to \frac12^-$ transitions the form factor decomposition reads
\begin{eqnarray}
\label{eq:1212}
\big\langle \Lambda_c^{*1/2^-}, r'\ \vec{P}^{\,\prime}\left|\,
J_{bc}^\mu(0)\, \right| \Lambda_b, r\ \vec{P}
\big\rangle& =& {\bar u}^{\Lambda_c^*}(\vec{P}^{\,\prime}, r')\Big\{
\gamma^\mu\left[F_1\gamma_5- G_1\right]+ v^\mu\left[F_2\gamma_5
-G_2\right]+v'^\mu\left[F_3\gamma_5- G_3
\right]\Big\}u^{\Lambda_b}(\vec{P}, r \,) \label{eq:def_ff}
\end{eqnarray}
 The $u_{r}$ are dimensionless Dirac spinors (${\bar u}_{r'} u_r
 = \delta_{r r'}$),  $v^\mu$, $v'^\mu $ are the four velocities of
 the initial and final baryons and  the three vector (axial)  $F_1,\,F_2,\,F_3$ ($G_1,\,G_2,\,G_3$) form factors are 
 functions of $\omega$ or equivalently of $q^2$. For $\frac12^+ \to \frac32^-$ decays we write
\begin{eqnarray}
\label{eq:1232}
&&\big\langle \Lambda_c^{*3/2^-},r'\vec P'\,|\,J_{bc}^\mu(0) \,|\,\Lambda_b,r\,
\vec P\,\big\rangle  = 
~\bar{u}^{\Lambda_c^*}_{\alpha}(\vec{P}', r')\,\Gamma^{\alpha\mu}\,
u^{\Lambda_b}(\vec{P}, r\,)
\nonumber\\
\Gamma^{\alpha\mu}&=& v^\alpha\Big\{ \gamma^\mu\left[l_{V_1}-l_{A_1}\gamma_5\right] + v^\mu\left[l_{V_2}-l_{A_2}\gamma_5\right]+ v'^\mu\left[l_{V_3}-l_{A_3}\gamma_5\right]\Big\}+ g^{\alpha\mu} \left[l_{V_4}-l_{A_4}\gamma_5\right]
\end{eqnarray}
Here $u^{\Lambda_c^*}_{\alpha\,r'}$ is the Rarita-Schwinger spinor of the final spin
3/2 baryon normalized such that $(\bar u_{\alpha\,r'}^{B'})
u^{B'\,\alpha}_r = -\,\delta_{rr'}$, and we have four vector
($l_{V_{1,2,3,4}}(\omega)$) and four axial ($l_{A_{1,2,3,4}}(\omega)$) form
factors.
\subsection{Infinite heavy quark mass limit}
\label{sec:ihqml}

The single heavy baryon and heavy quark velocities are equal in the $m_Q\to \infty$ limit. The heavy baryon can be viewed as a freely
propagating point-like color source (the heavy quark), dressed by strongly interacting
brown muck bearing appropriate color, flavor, baryon number, energy,
angular momentum and parity to make up the observed physical state. Since an infinitely massive heavy
quark does not recoil from the emission and absorption of soft $(E\sim \Lambda_{\rm QCD})$ gluons,
and since chromomagnetic interactions of such a quark are suppressed as $1/m_Q$, neither its mass (flavor) nor its spin affect 
the state of the light degrees of freedom. This results in a remarkable simplification of the description
of transitions in which a hadron containing a heavy quark, with velocity $v^\mu$, decays
into another hadron containing a heavy quark of a different flavor. To the heavy
quark, this looks like a free decay (up to pertubative QCD corrections), in which
the light dressing plays no role. The brown muck, on the other hand, knows only
that its point-like source of color is now recoiling at a new velocity $v'^{\mu}$, and it must
rearrange itself about it in some configuration~\cite{Falk:1991nq}.  Hence, in the $m_Q\to \infty$ limit, the weak matrix elements must become invariant under independent spin rotations of the $c$ and $b$ quarks. This is easily shown in the brick wall frame ($\vec{v}=-\vec{v}\,'$, $v^0=v'^0$) by quantizing the angular momentum of the {\it ldof} (brown muck) about the spatial axis defined by $\vec{v}$. It follows that neither the initial and final heavy baryons, nor the  $c$ and $b$ quarks have 
orbital angular momentum about this decay axis. Thus, in the
Isgur-Wise (IW) limit, the spins of $c$ and $b$ are decoupled from the light quanta, and the component of the {\it ldof} total angular momentum along the decay axis is conserved~\cite{Politzer:1990ps, Falk:1991nq}. 

This large spin invariance in the $m_Q\to \infty$ limit leads to considerable simplifications~\cite{Isgur:1990pm, Georgi:1990cx}. In particular, the semileptonic decay of the ground state $\Lambda_b$  into either $\Lambda_c^*$  in the $j_q^P= 1^-$ heavy doublet is described by an universal form-factor~\cite{Isgur:1991wr}. In this limit,  the  bottom quark carries all of the angular momentum of the $\Lambda_b$,  where the {\it ldof} are coupled to $j_q^P=0^+$. Within the tensor representation of the heavy baryon states~\cite{Falk:1991nq}, the $\Lambda_b$ is accounted 
by a Dirac  spinor $u_b(v)$, with $v$ the velocity of the $\Lambda_b$ (and of its heavy point-like constituent), satisfying the subsidiary condition  
$\Slash{v}u_b(v)=u_b(v)$. The charm $j_q^P= 1^-$ doublet of baryons, with four velocity $v'$, are represented by the multiplet-spinor ${\cal U}^\mu_c(v')$ 
\begin{equation}
{\cal U}^\mu_c(v') = u^{3/2\mu}_c(v') + \frac{1}{\sqrt3}(\gamma^\mu + v'^\mu)\gamma_5 u^{1/2}_c(v') \label{eq:genU}
\end{equation}
where the Dirac  $u^{1/2}_c(v')$ and the Rarita-Schwinger spinors $u^{3/2\mu}_c(v')$ stand for the spin 1/2 and spin 3/2 members of this doublet, respectively. The multiplet-spinor in Eq.~(\ref{eq:genU}) satisfies $\Slash{v}'{\cal U}^\mu_c(v')={\cal U}^\mu_c(v')$, and $v'_\mu{\cal U}^\mu_c(v')=0$. Note also that $\gamma_\mu u^{3/2\mu}_c=0$~\cite{Falk:1991nq}.

Under a Lorentz transformation, $\Lambda$, and 
$b$ and $c$ quark spin transformations $\widehat S_b$ and $\widehat S_c$, the above spinor wave functions
transform as $S(\Lambda)\,u_b,\, \Lambda^\mu_\nu S(\Lambda)\, {\cal U}^\nu_c$ and $\widehat S_b\, u_b$ and $\widehat S_c \,{\cal U}^\mu_c$, respectively, with $S(\Lambda)=\exp\{-i\sigma_{\mu\nu}S^{\mu\nu}/4\}$, the usual spinor representation. Note that $\widehat S_b$ and $\widehat S_c$ are also of the form $S(\hat\Lambda)$, but with $\hat\Lambda$ restricted to spatial rotations and affecting only to the heavy quark spinor.

In addition in the $m_Q\to \infty$ limit, under heavy  quark spin rotations, the $b\to c$ flavor changing current transforms as $J^\mu_{bc} \to \widehat S_c J^\mu_{bc} \widehat S_b^\dagger$. With all these ingredients, the most general form for the matrix element respecting 
HQSS is~\cite{Isgur:1991wr,Leibovich:1997az}
\begin{equation}
 \big\langle \Lambda_c^{*1/2^-,\,3/2^-}; j_q^P=1^-\left|\,
J_{bc}^\mu(0)\, \right| \Lambda_b\big\rangle = \sigma(\omega) v_\lambda\, \bar{\cal U}^\lambda_c(v') \gamma^\mu (1-\gamma_5) u_b(v) + {\cal O}(1/m_{b,c})\,,\label{eq:defIWinfty}
\end{equation}
Here $\sigma(\omega)$ is the (real) dimensionless leading IW function for the transition to this excited doublet. It follows $\sqrt3\,F_1/(\omega-1)=\sqrt3\,G_1/(\omega+1)=-\sqrt3\,F_2/2=-\sqrt3\,G_2/2=l_{V_1}=l_{A_1}=\sigma$ and $F_3=G_3=l_{V_{2,3,4}}=l_{A_{2,3,4}}=0$. 
In Ref.~\cite{Leibovich:1997az},   $\sigma(\omega)$ was predicted in the large $N_c$ limit,
\begin{equation}
 \sigma(\omega) = 1.2 \left[ 1-1.4(\omega-1)\right] \label{eq:sig-num}
\end{equation}
where subleading $1/N_c$ corrections are neglected.

The matrix element in Eq.~(\ref{eq:defIWinfty}) vanishes at zero recoil, where $v=v'$, and it trivially leads to\footnote{The sum over the initial and final polarizations in the definition of the hadronic tensor in Eq.~(\ref{eq:wmunu}) can be written as trace in the Dirac space, with  the help of the spin 1/2  and 3/2  projectors. These latter operators are $u(v)\,\bar u(v) = (1+\Slash{v})/2$ and $u^\rho(v)\,\bar u^\lambda(v) =\left(-g^{\rho\lambda}+v^\rho\,v^\lambda+ (\gamma^\rho+v^\rho)(\gamma^\lambda-v^\lambda)/3 \right) (1+\Slash{v})/2$. } 
\begin{equation}
 {\cal H}^{\alpha\beta}_{3/2^-}[j_q^P=1^-] = 2 {\cal H}^{\alpha\beta}_{1/2^-} [j_q^P=1^-]   = \frac{2\sigma^2(\omega)}{3} (\omega^2-1)\, \left( v^\alpha v'^\beta+v^\beta v'^\alpha-\omega g^{\alpha\beta} -i  \epsilon^{\alpha\beta\rho\sigma}v_\rho v'_\sigma\right) + {\cal O}(1/m_{b,c}) \label{eq:me-unomenos}
\end{equation}
the antisymmetric term does not contribute to $d\Gamma/d\omega$ since the leptonic tensor, after integrating
in the lepton momenta, becomes symmetric. Thus in the $m_Q\to +\infty$ limit, $d\Gamma_{\Lambda_c^{*3/2}}/d\omega = 2 d\Gamma_{\Lambda_c^{*1/2}}/d\omega$ since both members  of the $j^P_q= 1^-$ doublet are degenerate. Furthermore, one easily deduces that $\Lambda_b$ decays to excited $\Lambda_c^{*3/2^-}$ with helicity $\pm 3/2 $ are forbidden by HQSS in the IW limit, since the component of the {\it ldof} total angular momentum along the decay axis is conserved, and equal to zero. 

On the other hand, for the ground-state $\Lambda_b$ transition to the $J^P=1/2^-$ charmed baryon with $j_Q^P=0^-$ {\it ldof}, one can use for the latter a spinor $u_c(v')$, but the form-factors must be pseudoscalar and therefore involve a Levi-Civita tensor~\cite{Mannel:1990vg}. At leading order in the $1/m_Q$ expansion, there are not enough vectors available to contract with the indices of the epsilon tensor so these unnatural\footnote{A semileptonic baryonic transition is unnatural  if it involves transitions between tensor ($0^+, 1^-, 2^+, \cdots$) to pseudo-tensor ($0^-, 1^+, 2^-, \cdots$), or vice-versa, $j_q^P$ {\it ldof} quantum numbers.  } parity matrix elements vanish~\cite{Roberts:1992xm,Leibovich:1997az}. 

A different way to understand why the $\Lambda_b[1/2^+, j_q^P=0^+] \to \Lambda_c^*[1/2^-, j_q^P=0^-]$ is forbidden in the IW limit is adopting the picture introduced in Refs.~\cite{Liang:2016exm, Liang:2016ydj}. In the heavy-quark limit, 
the weak transition occurs on the $b$ quark, which turns into a $c$ quark and a $W^-$ boson, as shown in the left panel of Fig.~\ref{fig:SL}. Since we will have a $1/2^-$ or $3/2^-$ state at the end,  and the $u, d$ quarks are spectators, remaining in a $0^+$ spin-parity configuration, 
the final charm quark  must carry negative parity and hence must be in an $L = 1$ level. This corresponds to an orbital angular momentum excitation 
between the heavy quark and the isoscalar $u,d$ diquark as a whole, which maintains the same spin-parity quantum numbers, $0^+$, as in the initial $\Lambda_b$, leading to a non-zero {\it ldof} wave-function overlap. Within this picture, the total angular momentum and parity of the light subsystem will be $j_q^P=1^-\,[= 0^+\otimes (L=1)]$, and the transition will be described by the matrix element in Eq.~(\ref{eq:defIWinfty}), that will go through $P-$wave, giving rise to the $(\omega^2-1)$  factor in 
Eq.~(\ref{eq:me-unomenos}). In sharp contrast, the $(j_q^P=0^-,J^P=1/2^-)$ final baryon contains a $P-$wave excitation inside the brown muck and a realignment of the light quarks spins to construct a spin triplet state. That requires going beyond the spectator 
approximation of Fig.~\ref{fig:SL}, involving dynamical changes in the QCD dressing of the heavy baryon during the transition, which are $1/m_Q-$ suppressed. Thus in the heavy quark limit,  the initial and final {\it ldof} overlap for the unnatural $0^+\to 0^-$ transition vanishes. It would be parametrized by a pseudoscalar form-factor, involving  the Levi-Civita tensor. As mentioned above,  at leading order in the $1/m_Q$ expansion, there are not enough vectors available to contract with the indices of the epsilon tensor.

\subsection{${\cal O}(\Lambda_{\rm QCD}/m_c)$ corrections}
\label{sec:invmc}
Corrections of order $1/m_Q$ to $d\Gamma(\Lambda_b\to \Lambda_c^{*3/2}[j^P_q=1^-])/d\omega$ and  $d\Gamma(\Lambda_b\to \Lambda_c^{*1/2}[j^P_q=1^-])/d\omega$ distributions were studied in \cite{Leibovich:1997az} and  shown to be quite large, specially in the  $J^P=1/2^-$ case (see Fig. 1 of that reference).

Neglecting ${\cal O}(\Lambda_{\rm QCD}/m_b)$ terms, this is to say keeping still the invariance of the weak matrix element under arbitrary $b-$quark spin rotations, the  general forms of the semileptonic matrix elements are
\begin{eqnarray}
 \big\langle \Lambda_c^{*1/2^-}\left|\,
J_{bc}^\mu(0)\, \right| \Lambda_b\big\rangle &=& \frac1{\sqrt3} \bar u_c(v')\left[(\Slash{v}-\omega)\Delta_1-\Delta_2\right] \gamma^\mu (1-\gamma_5) u_b(v) + {\cal O}(1/m_{b})\,,\label{eq:defIWinfty-1mc12} \\
\big\langle \Lambda_c^{*3/2^-}\left|\,
J_{bc}^\mu(0)\, \right| \Lambda_b\big\rangle &=&  \bar u_c^\lambda(v')v_\lambda\left[\Omega_1-(\Slash{v}-\omega)\Omega_2\right] \gamma^\mu (1-\gamma_5) u_b(v) + {\cal O}(1/m_{b})\,,\label{eq:defIWinfty-1mc32} 
\end{eqnarray}
where $\Delta_{1,2}$ and $\Omega_{1,2}$  are form factors function of $\omega$ that are used to construct independent linear combinations of the identity and $\Slash{v}$ matrices. For semileptonic transitions to $\Lambda_c^{*1/2^-}$, we find $\sqrt3 F_1= (\omega-1)\Delta_1+\Delta_2$, 
$\sqrt{3} G_1=(\omega+1)\Delta_1+\Delta_2$, $F_2=G_2=-2\Delta_1/\sqrt3$ and $F_3=G_3=0$. Similarly for $\Lambda_c^{*3/2^-}$, we find $l_{V_1}= \Omega_1+(\omega+1)\,\Omega_2$, $l_{A_1}= \Omega_1+(\omega-1)\,\Omega_2$, $l_{V_2}= l_{A_2}= -2\,\Omega_2$ and $l_{V_{3,4}}= l_{A_{3,4}}=0$.

If $\Delta_1= \Omega_1= \sigma$ and $\Delta_2=\Omega_2=0$, the IW limit of Eq.~(\ref{eq:me-unomenos}) is recovered for transitions to $\Lambda_c^{*3/2^-,\, 1/2^-}\,[j_q^P=1^-]$ states\footnote{Note that for the $1/2^-$ member of the $j_q^P=1^-$ multiplet, we have 
$v_\lambda \left[(\gamma^\lambda + v'^\lambda\right)\gamma_5 u^{1/2}_c(v')]^\dagger \gamma^0 = \bar u_c^{1/2}(v')\left(\Slash{v}-\omega \right)\gamma_5$ and $\gamma_5\gamma^\mu (1-\gamma_5)=\gamma^\mu (1-\gamma_5) $.}.

The  differential decay widths deduced from the general matrix elements of Eqs.~(\ref{eq:defIWinfty-1mc12}) and ~(\ref{eq:defIWinfty-1mc32}) are  given by
\begin{eqnarray}
\frac{d\Gamma[\Lambda_b\to \Lambda_c^*(J^P)]}{d\omega}&=& C_J \frac{8\,\Gamma_0}{3}\left(\frac{M_{\Lambda_c^*}}{M_{\Lambda_b}}\right)^3\left(1-\frac{m^2_\ell}{q^2}\right)^2(\omega^2-1)^J\,
\left\{ \alpha_J^2\left[3\omega \frac{q^2+m_\ell^2}{M_{\Lambda_b}^2} + 2\frac{M_{\Lambda_c^*}}{M_{\Lambda_b}}(\omega^2-1)\left(1+\frac{2m^2_\ell}{q^2}\right)\right]\right.\nonumber \\
&&\left. +2\,(\omega^2-1)\,\left[\alpha_1(\omega)\alpha_2(\omega)\right]_J \left[\frac{2q^2+ m_\ell^2}{M_{\Lambda_b}^2} +\left(1-\frac{M^2_{\Lambda_c^*}}{M_{\Lambda_b}^2}\right)\left(1+\frac{2m^2_\ell}{q^2}\right)\right]\right\}+ {\cal O}(1/m_{b})\,,\label{eq:gamma_diff}
\end{eqnarray}
with $J^P= 1/2^-, \, 3/2^-$, $C_J = \left(2J+1\right)$ and
\begin{eqnarray}
 \alpha_{J=1/2}^2(\omega) &=& \Delta_2^2(\omega)+(\omega^2-1)\Delta_1^2(\omega), \qquad \left. \alpha_{1}(\omega)\,\alpha_{2}(\omega)\,\right|_{J=1/2} =\Delta_1(\omega)\,\,\Delta_2 (\omega)\,\\
 \alpha_{J=3/2}^2(\omega) &=& \Omega_1^2(\omega)+(\omega^2-1)\Omega_2^2(\omega), \qquad \left. \alpha_{1}(\omega)\,\alpha_{2}(\omega)\,\right|_{J=3/2} =\Omega_1(\omega)\,\,\Omega_2(\omega) 
\end{eqnarray}

At order ${\cal O}(\Lambda_{\rm QCD}/m_Q)$, there are corrections originating from the matching of the $b\to c$ 
flavor changing current onto the heavy quark effective
theory and from order $\Lambda_{\rm QCD}/m_Q$ corrections to the effective Lagrangian~\cite{Georgi:1990ei,Neubert:1993mb, Roberts:1992xm, Leibovich:1997az, Mannel:1991ii}. Following the discussion of Ref.~\cite{Leibovich:1997az}, for $\Lambda_b$ decays, they have a quite different physiognomy depending on the total angular momentum and parity of the {\it ldof} in the daughter charm excited baryon. In particular, 
\begin{itemize}
 \item $j_q^P= 1^-$: Neglecting $1/m_b$ corrections and QCD short-range logarithms~\cite{Leibovich:1997az},
 \begin{eqnarray}
   \Delta_1(\omega) &=& \sigma(\omega) + \frac{1}{2m_c}\left(\phi_{\rm kin}^{(c)}(\omega) - 2 \phi_{\rm mag}^{(c)}(\omega)\right),\quad \Delta_2(\omega)= 
   \frac{1}{2m_c} \left( 3(\omega \bar\Lambda'-\bar\Lambda)\sigma(\omega)+2(1-\omega^2)\sigma_1(\omega) \right) \label{eq:Deltas} \\
   \Omega_1(\omega) &=& \sigma(\omega) + \frac{1}{2m_c}\left(\phi_{\rm kin}^{(c)}(\omega) + \phi_{\rm mag}^{(c)}(\omega)\right),\quad 
   \Omega_2(\omega)= 
   \frac{\sigma_1(\omega)}{2m_c} \label{eq:Omegas}
  \end{eqnarray}
with $m_c\sim 1.4$ GeV, the charm quark mass, and $\bar\Lambda\sim 0.8$ GeV [$\bar\Lambda'\sim (1 \pm 0.1$ GeV)] 
the energy of the {\it ldof} in the $m_Q\to \infty$ limit in the $\Lambda_b$ [$\Lambda_c^*\,(j_q^P=1^-)$] baryon. The $\sigma_1(\omega)$ form-factor  
determines, together with $\bar \Lambda$ and $\bar \Lambda'$,  the $1/m_c$ corrections stemming from the matching of the QCD and effective theory currents. This sub-leading IW function is unknown and in Ref.~\cite{Leibovich:1997az},  it was  varied   in the range  
$\pm 1.2 \left[ 1-1.6(\omega-1)\right]$ GeV. In addition, $\phi_{\rm kin}^{(c)}$ and $\phi_{\rm mag}^{(c)}$ account for 
the time ordered product of the dimension-five kinetic energy and chromomagnetic operators in the effective Lagrangian.  The chromomagnetic term
is neglected in \cite{Leibovich:1997az}, because it is argued that it should be small relative to $\Lambda_{\rm QCD}$. In addition,  the kinetic energy correction  is estimated in the large $N_c$ limit,  $\phi_{\rm kin}^{(c)} = -\frac{\bar \Lambda}{8}\sqrt{\frac{\bar \Lambda^3}{\kappa}}\left(\omega^2-1\right)\sigma(\omega)$, with $\kappa=(0.411\,{\rm GeV})^3$~\cite{Leibovich:1997az}. 

The Eqs.~(\ref{eq:Deltas}) and (\ref{eq:Omegas}) can be re-derived  from
\begin{equation}
 \big\langle \Lambda_c^{*1/2^-,\,3/2^-}; j_q^P=1^-\left|\,
J_{bc}^\mu(0)\, \right| \Lambda_b\big\rangle = \bar{\cal U}^\lambda_c(v')\, \left\{v_\lambda[\beta_1+(\omega-\Slash{v})\beta_2]+\gamma_\lambda \beta_3/3\right\} \gamma^\mu (1-\gamma_5) u_b(v) + {\cal O}(1/m_{b})\,,\label{eq:defIWinfty.gral}
\end{equation}
where the ${\cal O}(1/m_c)$ $\beta_2$ and $\beta_3$ form-factors and the sub-leading term of $\beta_1$ depend on $J$. Thus, we have
 \begin{eqnarray}
   \beta_1(\omega)|_J &=& \sigma(\omega) + \frac{1}{2m_c}\left(\phi_{\rm kin}^{(c)}(\omega) +c_J \phi_{\rm mag}^{(c)}(\omega)\right) \nonumber \\
   \beta_2(\omega)|_J &=& \frac{c_J}{2m_c}\, \sigma_1(\omega)\, ,\quad   
   \beta_3(\omega) = 3\,\frac{(\omega \bar\Lambda'-\bar\Lambda)}{2m_c}\,\sigma(\omega) \label{eq:betas}
   \end{eqnarray}
with $c_{J=1/2}= -2$ and    $c_{J=3/2}= 1$, which correspond to the eigenvalues of the operator $2\, \vec{S}_c \cdot \vec{j}_q$
\begin{equation}
c_J= J\,(J+1)-\frac12\,(\frac12+1)-1\,(1+1)\,, \label{eq:cjs}
\end{equation}
for $j_q= 1$ and $S_c=1/2$, and 
 \begin{equation}
   \Omega_1 = \beta_1(\omega)|_{J=3/2} \,, \quad \Delta_1 = \beta_1(\omega)|_{J=1/2}, \quad \Omega_2 = \beta_2(\omega)|_{J=3/2}\,,\quad  \Delta_2 = \beta_3(\omega)+\beta_2(\omega)|_{J=1/2}. \label{eq:betas-bis}
   \end{equation}
The $1/m_b$ contributions, not taken into account, are much smaller than the theoretical uncertainties induced by the errors
on $(\bar \Lambda-\bar\Lambda')$ and the $\sigma_1(\omega)$ form-factor. Hence, the form-factors of Eqs.~(\ref{eq:Deltas}) and (\ref{eq:Omegas}) provide an excellent approximation to the results reported in Ref.~\cite{Leibovich:1997az}.

Two final remarks to conclude this discussion:  i) The $(\omega\bar\Lambda'- \bar\Lambda)$ difference in $\Delta_2$ [$\gamma_\lambda$ form-factor in Eq.~(\ref{eq:defIWinfty.gral})] provides a $S-$wave $W^-\Lambda_c^* (1/2^-)$ term that should scale as $\sqrt{\omega^2-1}$, and hence should dominate this differential rate at zero recoil. ii) The kinetic operator correction is the only $1/m_c$ term that does not break HQSS.

\item $j_q^P= 0^-$: For the case of this unnatural transition, the matrix elements of the $1/m_Q$  current and kinetic energy operator corrections are zero for the same reason that the leading form factor vanished~\cite{Leibovich:1997az}. The time ordered products involving the chromomagnetic 
operator lead to non-zero contributions, which however vanish at zero recoil~\cite{Leibovich:1997az} and can be cast in a $\Delta_1-$type form factor. At order $1/m_Q$ the corresponding $\Delta_2$ form-factor is zero. 

From the above results, we conclude that the $\Lambda_b$ semileptonic decay to a $J^P=1/2^--$daughter charm excited baryon with a $j_q^P= 0^-$ 
{\it ldof}--configuration can be visible only if HQSS is severely broken and higher $\left(1/m_Q\right)^n$ corrections are sizable. 

\end{itemize}
\subsection{Decays to molecular $\Lambda_c^{\rm MOL}$ states}
\label{sec:molecular-decays}
Following the spectator image of Fig.~\ref{fig:SL}, the $c$ quark created in the weak transition  
must carry negative parity and hence must be in a relative $P-$wave. The parity and total angular momentum of the final resonance are those of the intermediate system before hadronization. Since the molecular $\Lambda_c^{\rm MOL}$ states come from 
meson-baryon interaction in our picture, we must hadronize the final state including a  $q\bar q$ pair with the quantum numbers
of the vacuum ($^{2S+1}L_J=^3P_0$). This is done following the work of Refs.~\cite{Liang:2016exm, Liang:2016ydj}, and thus we include $u\bar u+d\bar d + s\bar s$ as in 
the right panel of Fig.~\ref{fig:SL}. The $c$ quark must be involved in the hadronization, because it is originally in an $L = 1$ state, but after the
hadronization produces the $D^{(*)}N$ state, and the $c$ quark in the $D^{(*)}$ meson is in an $L = 0$ state. Neglecting hidden-strange contributions, the hadronization results in isoscalar $S-$wave $DN$ and $D^*N$ pairs, but does not produce  $\pi\Sigma_c^{(*)}$ states~\cite{Liang:2016exm, Liang:2016ydj}.

The production of $J^P=1/2^-, 3/2^-$ resonances ($R_J$) is done after the created $DN$ and $D^*N$ in the first step couple into the resonance, as shown in the right panel of Fig.~\ref{fig:SL}. The transition matrix, $t_{R_J}$, for such
mechanism leads to
\begin{eqnarray}
 \overline{\sum}\sum |t_{R_J}|^2 &=&  \sum_M {\cal C}(\frac12 1 J|  M 0 M)^2 \, |\varphi(\omega)\,|^2 \, \left | C_J^{DN}\,g_{R_J}^{DN}\, G_{DN} + C_J^{D^*N}\,g_{R_J}^{D^*N}\, G_{D^*N}  \right|^2, \label{eq:begin-mol}
\end{eqnarray}
where the sums are over the spins of the initial and final particles, and the bar over the sum denotes the average over initial spins. The Clebsch-Gordan coefficient accounts for the coupling of spin and orbital angular momentum of the $c-$quark to the total angular momentum $J$ of the 
intermediate system, composed by the charm quark and the spectator isoscalar $0^+$ $ud$ diquark (see Fig.~\ref{fig:SL}). Because angular momentum conservation, the spin of the resonance, produced after hadronization and meson-baryon re-scattering,  will be $J$ as well. The important point is that the third component of the orbital angular momentum of the $c-$quark must be zero~\cite{Politzer:1990ps, Falk:1991nq} (see also the 
discussion at the beginning of Subsec.~\ref{sec:ihqml}). Let us note for future purposes that ${\cal C}(\frac12 1 \frac32|  M 0 M)^2\,/\, {\cal C}(\frac12 1 \frac12 | M 0 M)^2 = 2, \, M= \pm 1/2$. 

The function $\varphi(\omega)$ accounts for some $\omega$ dependences induced by the hadronization process and 
by the matrix element between the initial $S-$wave $b-$quark,  the outgoing $W-$plane wave and the $P-$wave $c-$quark created in the intermediate hadronic state. This latter factor should scale like  $|\vec{q}\,| \propto \sqrt{\omega^2-1}$ close to zero recoil~\cite{Liang:2016exm, Liang:2016ydj}. In the heavy quark limit assumed in the mechanism depicted in Fig.~\ref{fig:SL}, one expects $\varphi(\omega)$ to be independent of the  angular momentum, $J$, of the final resonance.

The $C_J^{D^{(*)}N}$ coefficients  account for different overlaps between $DN$ and $D^*N$ $S-$wave pairs and the intermediate hadronic state, whose wave-function is determined by a excitation among the  heavy quark and the brown muck ({\it ldof}) as a whole. This is a $\lambda-$excited state in the framework of CQM's, 
and it has $j_q^P=1^-$  quantum-numbers for the brown muck. The values of $C_J^{D^{(*)}N}$ can be readily obtained from Eq.~(\ref{eq:relation}),
\begin{equation}
 C_J^{D^{(*)}N} = \langle D^{(*)} N; J\,|\,S_Q^P=1/2^+\, j_q^P=1^-; J \rangle_2
\end{equation}
Finally, $G_{D^{(*)}N}$ is the loop function for the $D^{(*)}N$ propagation\footnote{We are assuming that $G_{D^*N}$ is the same both for $J=1/2$ and $J=3/2$. This is correct as long as the renormalization of the UV divergences of this loop function does not depend on the angular momentum, as in the 
${\rm SU(6)}_{\rm lsf} \times {\rm SU(2)}_{\rm HQSS}$  and ELHG models of Refs.~\cite{GarciaRecio:2008dp, Romanets:2012hm} and \cite{Liang:2014kra}, respectively.} and $ g_{R_J}^{D^{(*)}N}$ is the dimensionless coupling of the resonance $R_J$ to the $D^{(*)} N$
channel in isospin zero.  They are defined for instance in Eqs.~(15) and (18) of Ref.~\cite{GarciaRecio:2008dp}, and we compute them at the resonance position in the complex plane. Note that the couplings  
$g_{R_J}^{D^{(*)}N}$, obtained from the residues of the coupled-channels meson-baryon $T-$matrix, contain effects from intermediate $\pi\Sigma_c^{(*)}$ loops.

With all these ingredients close to zero recoil, we find 
\begin{equation}
\left. \frac{d\Gamma/d\omega[\Lambda_b\to \Lambda_c^*(1/2^-)]}{d\Gamma/d\omega[\Lambda_b\to \Lambda_c^*(3/2^-)]}\right|_{\rm MOL} = \frac12 \,\frac{\left | \frac{\sqrt{3}}{2}\,g_{R_{J=1/2}}^{DN}\, G_{DN} + \frac12 \,g_{R_{J=1/2}}^{D^*N}\, G_{D^*N}  \right|^2}{\left |g_{R_{J=3/2}}^{D^*N}\, G_{D^*N}\right|^2} \label{eq:decaymol}
\end{equation}
where the factor $1/2$ comes from the ratio of Clebsch-Gordan coefficients. In this way, we recover the main result of  Ref.~\cite{Liang:2016exm}. It shows that the above ratio of differential decay widths is very sensitive to the couplings of 
the  $\Lambda_c^*$ resonances to the $DN$ and $D^*N$ channels. We could expect Eq.~(\ref{eq:decaymol}) to hold also in good approximation for the ratio of integrated rates since the available phase space is quite small

In the infinite heavy quark mass limit,  
the degeneracy of the $D$ and $D^*$ masses implies $G_{DN}=G_{D^*N}$. In addition for $1^-$ and $0^-$ {\it ldof} quantum numbers,   
the couplings of $DN$ and $D^*N$ to $\Lambda_c^*$ are related
\begin{eqnarray}
j_q^P= 1^- &\Rightarrow & \frac{2}{\sqrt{3}}\, g_{\Lambda_c^*(1/2^-)}^{DN}=2g_{\Lambda_c^*(1/2^-)}^{D^*N}=g_{\Lambda_c^*(3/2^-)}^{D^*N} \\
j_q^P= 0^- &\Rightarrow &
\sqrt{3}\,g_{\Lambda_c^*(1/2^-)}^{DN}=-g_{\Lambda_c^*(1/2^-)}^{D^*N}\,,
\end{eqnarray}
as inferred from Eq.~(\ref{eq:relation}). Hence, we re-obtain the 
$ m_Q\to \infty $ results of Subsec.~\ref{sec:ihqml},  
\begin{equation}
 \frac{d\Gamma/d\omega[\Lambda_b\to \Lambda_c^*(1/2^-)]_{j^p_q =1^-}}{d\Gamma/d\omega[\Lambda_b\to \Lambda_c^*(3/2^-)]_{j^p_q =1^-}} = \frac12,\qquad  \frac{d\Gamma/d\omega[\Lambda_b\to \Lambda_c^*(1/2^-)]_{j^p_q =0^-}}{d\Gamma/d\omega[\Lambda_b\to \Lambda_c^*(3/2^-)]_{j^p_q =1^-}} = 0
\end{equation}
For molecular states, we might have deviations from the above IW limit predictions, and in particular visible widths for a charm $J^P=1/2^-$ excited baryon with  significant $0^-$ {\it ldof} components. This could  happen if  the meson-baryon interactions, which generate the molecular state, induce important  $\left(1/m_Q\right)^n$  corrections,  bigger than would be expected from the discussion in  Subsec.~\ref{sec:invmc}.
\section{$\Lambda_b \to \Lambda_c^* \pi^-$ decay}
\label{sec:pidecay}
Looking again at the diagram depicted in the left panel of Fig.~\ref{fig:SL}, the $\Lambda_b \to \Lambda_c^* \pi^-$ decay could proceed through the mechanism
of external emission~\cite{Chau:1982da}, where the gauge $W^-$ boson couples to $\pi^-$ instead of to the $(\ell^-\bar\nu_{\ell})$ lepton pair. This is the factorization approximation, which should be accurate for processes that involve a heavy hadron and multiple light
mesons in the final state, provided the light mesons are all highly collinear and energetic~\cite{Dugan:1990de}. Actually for 
$\Lambda_b \to \Lambda_c^* \pi^-$ decay, corrections are expected  
to be of the order $\Lambda_{\rm QCD}/E_\pi$, with $E_\pi$ the energy of the pion in the center of mass frame. There exist also some small strong coupling  
logarithmic corrections stemming from the matching of full QCD with the effective heavy quark theory. The $\Lambda_b \to \Lambda_c^* \pi^-$ width  is related to the differential decay rate $d\Gamma_{\rm sl}/d\omega$ at $q^2=m_\pi^2$ [$\omega=(M_{\Lambda_b}^2 + M_{\Lambda_c^*}^2-m_\pi^2)/2M_{\Lambda_b}M_{\Lambda_c^*}$] for the analogous semileptonic decay~\cite{Leibovich:1997az},
\begin{equation}
\Gamma_\pi[\Lambda_b \to \Lambda_c^* \pi^-] \propto |V_{ud}|^2\,f_\pi^2 \left.\frac{d\Gamma_{\rm sl}[\Lambda_b \to \Lambda_c^* e^-\bar{\nu}_e]}{d\omega}\right|_{q^2=m_\pi^2} 
\end{equation}
with $m_\pi $ and $f_\pi$, the pion mass and decay constant, respectively. In the case of decays into $\Lambda_c^*$ molecular states, we find again that the ratio of $gG$ factors of Eq.~(\ref{eq:decaymol}) provides an estimate for $\left. \frac{\Gamma_\pi[\Lambda_b\to \Lambda_c^*(1/2^-)]}{\Gamma_\pi[\Lambda_b\to \Lambda_c^*(3/2^-)]}\right|_{\rm MOL}$. However, the  kinematics now is significantly different to that of zero recoil. 
In the $M_{\Lambda_b}$ rest frame, the recoil three momentum  is of the order of 2.2 GeV, even larger than the charm quark mass. Hence, the approximation of neglecting the effects of   operators like $\vec{S}_c \cdot \vec{j}_q$ in the weak transition  becomes inappropriate, since 
factors proportional to $|\vec{q}\,|/m_c$ can be large in this kinematics [$(\omega^2-1)\sim 0.7$]. This type of operators couples the charm quark spin and the angular momentum of the {\it ldof} and induces dependences on $J$, the total angular momentum of the created hadron. In this situation, it can not be guarantied that the function $\varphi(\omega)$, introduced in Eq.~(\ref{eq:begin-mol}), is independent of $J$.  In fact, in Ref.~\cite{Liang:2016ydj} and in addition to the quotient of $gG$ coefficients, a factor $(\vec{q}^{\,2}+E_\pi^2)/E_\pi^2\sim 2$ was found that increased the value of the $\left. \frac{\Gamma_\pi[\Lambda_b\to \Lambda_c^*(1/2^-)]}{\Gamma_\pi[\Lambda_b\to \Lambda_c^*(3/2^-)]}\right|_{\rm MOL}$ ratio.  We will also use here this result,  with some precautions, and we will multiply by a factor of 2 the estimates for the latter ratio 
deduced from the $gG$ factors.
\section{Results}
\label{sec:results}
\subsection{Semileptonic ($\mu^- \bar \nu_\mu$ or $e^- \bar \nu_e$) and pion $\Lambda_b\to \Lambda_c^*$ decays}
\label{sec:results-massless}
\begin{table*}[t!]
\begin{tabular}{c|cccc|c}
& IW$_\infty$ & IW$_{{\cal O}(1/m_Q)}$ & ${\rm SU(6)}_{\rm lsf} \times {\rm SU(2)}_{\rm HQSS}$ & ELHG & RPP  \\\hline
 $\Gamma_{\rm sl}[\Lambda_b \to \Lambda_{c\, (n)}(2595)]/\Gamma_{\rm sl}[\Lambda_b \to \Lambda_c(2625)]$ &$0.5$ & $1.4^{+1.7}_{-1.0}$&0.14 & 0.39\,--\,0.48&$0.6^{+0.4}_{-0.3}$ \\
 $\Gamma_{\rm sl}[\Lambda_b \to \Lambda_{c\, (b)}(2595)]/\Gamma_{\rm sl}[\Lambda_b \to \Lambda_c(2625)]$ &$-$ &$-$ &0.39 & $\sim 0.02$& $-$\\\hline
 $\Gamma_{\rm \pi}[\Lambda_b \to \Lambda_{c\, (n)}(2595)]/\Gamma_{\rm \pi}[\Lambda_b \to \Lambda_c(2625)]$ &$0.5$ & $1.4^{+3.3}_{-1.1}$ &0.14 -- 0.28 &0.76\,--\,0.91 & $1.0\pm 0.6$\\
 $\Gamma_{\rm \pi}[\Lambda_b \to \Lambda_{c\, (b)}(2595)]/\Gamma_{\rm \pi}[\Lambda_b \to \Lambda_c(2625)]$ &$-$ &$-$ &0.39 -- 0.78 &$\sim 0.02$  & $-$\\\hline
\end{tabular}
 \caption{ Ratios of semileptonic ($\mu^- \bar \nu_\mu$ or $e^- \bar \nu_e$) and pion $\Lambda_b$ decays into odd parity $J=1/2$ and $3/2$ charm baryons. We show predictions obtained from the molecular schemes of Refs.~\cite{GarciaRecio:2008dp,Romanets:2012hm} (${\rm SU(6)}_{\rm lsf} \times {\rm SU(2)}_{\rm HQSS}$) and  \cite{Liang:2014kra} (ELHG),  together with the $m_Q\to \infty$ limit (IW$_\infty$) ratios, and those found including the subleading corrections  [IW$_{{\cal O}(1/m_Q)}$] derived  in  Ref.~\cite{Leibovich:1997az} for the case of a $j_q^P=1^-$ HQSS doublet. The  ELHG results for 
 the narrow $\Lambda_c(2595)$ are taken from Refs.~\cite{Liang:2016exm, Liang:2016ydj}. In the case of molecular approaches, the  $gG$ factors that enter in Eq.~(\ref{eq:decaymol})  are compiled in Table~\ref{tab:gG}.
 The ranges quoted for the ${\rm SU(6)}_{\rm lsf} \times {\rm SU(2)}_{\rm HQSS}$ pion-mode ratios  account for the factor of two introduced at 
 the end of Sec.~\ref{sec:pidecay}, suggested by the findings of Ref.~\cite{Liang:2016ydj}.
 We also show in the last column experimental estimates for these ratios obtained from  branching fractions given in the RPP~\cite{Tanabashi:2018oca}. See the text for more details.} \label{tab:ratios} 
\end{table*}
\begin{table*}[t!]
    \begin{tabular}{c|rr|rr}
     & \multicolumn{2}{c|}{${\rm SU(6)}_{\rm lsf} \times {\rm SU(2)}_{\rm HQSS}$} &  \multicolumn{2}{c}{ELHG } \\
     & ~$g_{R_J}^{DN}\, G_{DN}$ &~ $g_{R_J}^{D^*N}\, G_{D^*N}$ &~ $g_{R_J}^{DN}\, G_{DN}$ & ~$g_{R_J}^{D^*N}\, G_{D^*N}$\\\hline
      $\Lambda_{c\, (n)}^{\rm MOL}(2595)$ &$-10.54 + 0.02\,i$  & $11.65 - 0.42\,i$~ & $\phantom{-}13.88-1.06\,i$ & $ 26.51 + 2.10\,i $ \\
      $\Lambda_{c\, (b)}^{\rm MOL}(2595)$ & $3.16 - 3.45\,i$  & $4.14 + 0.17\,i$~ &  $-0.68 + 3.13\,i$ & $-4.66 + 3.42\, i$\\
      $\Lambda_{c\,}^{\rm MOL}(2625)$     & $-$ & $-5.82 + 2.58\,i $~~& $-$ & $29.10$ \\\hline
    \end{tabular}
    \caption{ Values (MeV) of  the factors $g_{R_J}^{D^{(*)}N}\, G_{D^{(*)}N}$ 
    from Refs.~\cite{GarciaRecio:2008dp} (${\rm SU(6)}_{\rm lsf} \times {\rm SU(2)}_{\rm HQSS}$) and \cite{Liang:2014kra} (ELHG). 
    The signs of $g_{R_J=1/2}^{D^*N}\, G_{D^*N}$ are changed with respect to Ref.~\cite{Liang:2014kra}, as
discussed in \cite{Liang:2016ydj}. The values quoted  for the 
${\rm SU(6)}_{\rm lsf} \times {\rm SU(2)}_{\rm HQSS}$ $g_{R_J=1/2}^{D^*N}\,G_{D^*N}$  take into account the order meson-baryon used in this work to couple the spins (see 
footnote~\ref{foot:sign}).
\label{tab:gG}}
\end{table*}
In Table~\ref{tab:ratios}, we show results for the ratios of semileptonic ($\mu^- \bar \nu_\mu$ or $e^- \bar \nu_e$) and pion $\Lambda_b$ decays into odd parity $J=1/2$ and $3/2$ charm baryons, obtained within the molecular schemes of Refs.~\cite{GarciaRecio:2008dp,Romanets:2012hm} (${\rm SU(6)}_{\rm lsf} \times {\rm SU(2)}_{\rm HQSS}$) and  \cite{Liang:2014kra} (ELHG). 
As commented in Subsec.~\ref{sec:hadmol}, a double pole structure for the $\Lambda_c(2595)$ is found in these approaches, with clear similarities to 
the situation for the $\Lambda(1405)$, and hence we give results for both, the narrow $(n)$ and $(b)$ broad $\Lambda_c^{\rm MOL}(2595)$  states.  In Table~\ref{tab:ratios}, we also show experimental estimates for these ratios deduced from  
branching fractions given in the RPP~\cite{Tanabashi:2018oca}. We have considered that the 
reconstructed $\Lambda_c(2595)$ resonance observed in the  decays corresponds to the molecular narrow resonance. In addition, $m_Q\to \infty$ limit results   
(IW$_\infty$) and predictions obtained incorporating the subleading corrections (IW$_{{\cal O}(1/m_Q)}$) discussed in  Ref.~\cite{Leibovich:1997az} are also shown in Table~\ref{tab:ratios}. In this latter work, it is assumed that the $\Lambda_c(2595)$ and $\Lambda_c(2625)$ form the lowest-lying $j_q^P=1^-$ HQSS doublet, and the values quoted in the table follow mostly from Eqs.~(2.26) and (2.28) of that reference. To the error budget deduced from these equations, we have added in quadrature  the effects due to the uncertainty ($\pm 0.1 $ GeV) on the $\bar \Lambda'$ parameter in Eq.~(\ref{eq:Deltas}), which produces variations in the ratios of 
about 25\%--30\%~\cite{Leibovich:1997az}. The  errors on the IW$_{{\cal O}(1/m_Q)}$ ratios are largely dominated by the uncertainties on the subleading $\sigma_1$ form-factor.  It leads to opposite effects for $\Lambda_c(2595)$ or $\Lambda_c(2625)$ final states~\cite{Leibovich:1997az}, as can be inferred here from Eqs.~(\ref{eq:betas}) and (\ref{eq:cjs}). The biggest (smallest) $\Gamma^{\Lambda_{1/2}}_{\rm sl,\pi}/ \Gamma^{\Lambda_{3/2}}_{\rm sl,\pi}$ values correspond to $\sigma_1(1)=-1.2$ (+1.2)  GeV, while the central values are obtained for $\sigma_1(\omega)=0$. The 
$\Gamma^{\Lambda_{1/2}}_{\rm sl}$ rate, depending on $\sigma_1$, could be significantly enhanced (around a factor 2.5 for $\sigma_1=0$) compared to the infinite mass prediction ($\sim 0.020\,\Gamma_0)$, while $1/m_Q $ effects are much smaller for $\Gamma^{\Lambda_{3/2}}_{\rm sl}$. Predictions for the pion decay widths depend on $d\Gamma_{\rm sl}/d\omega$ at $q^2=m_\pi^2$, and turn out to be quite uncertain due to $\sigma_1$. We see that IW$_{{\cal O}(1/m_Q)}$ predictions and experimental estimates for the $\Gamma^{\Lambda_{1/2}}/ \Gamma^{\Lambda_{3/2}}$ ratios agree, within errors, for both semileptonic and pion $\Lambda_b$ decay modes. A certain tendency is observed in the central values, for which the theoretical estimations are greater than the experimental ones, in particular in the semileptonic mode. However, it would not be really significant due to the great uncertainties.

In what respects to the ELHG ratios for the narrow molecular $\Lambda_c(2595)$ state, we give in Table~\ref{tab:ratios} the ranges quoted in the original works of Refs.~\cite{Liang:2016exm, Liang:2016ydj}. The lowest ratios can be found using the $gG$ coefficients compiled in Table~\ref{tab:gG}, while the highest values account for corrections due to the contribution of hidden-strange ($D^{(*)}_s\Lambda$) channels in the hadronization.  Within the ELHG scheme the broad $\Lambda_c(2595)$ ratios  are negligible. This is because in this approach, the $J^P=3/2^-$ $\Lambda_c(2625)$ is a quasi-bound $D^*N$ state with a large coupling to this channel, whose absolute value is around five times bigger than that of the broad $\Lambda_c(2595)$ resonance to $D^*N$ or $DN$~\cite{Liang:2016ydj}. The narrow ELHG  $\Lambda_c(2595)$ molecule has $DN$ and $D^*N$ couplings (in absolute value) similar to $g_{\Lambda_c(2625)}^{D^*N}$, and its 
$\Gamma^{\Lambda_{1/2}}_{\rm sl,\pi}/ \Gamma^{\Lambda_{3/2}}_{\rm sl,\pi}$ ratios are larger and about 0.4 and 0.8, respectively,  compatible within errors with the experimental expectations. It should be also noted that after renormalization, the $DN$ loop function is almost a factor of two smaller than the $D^*N$ one, which produces a significant source of HQSS breaking in the ELHG approach of Ref.~\cite{Liang:2016ydj}. 

\begin{figure}[tbh]
\includegraphics[scale=0.9]{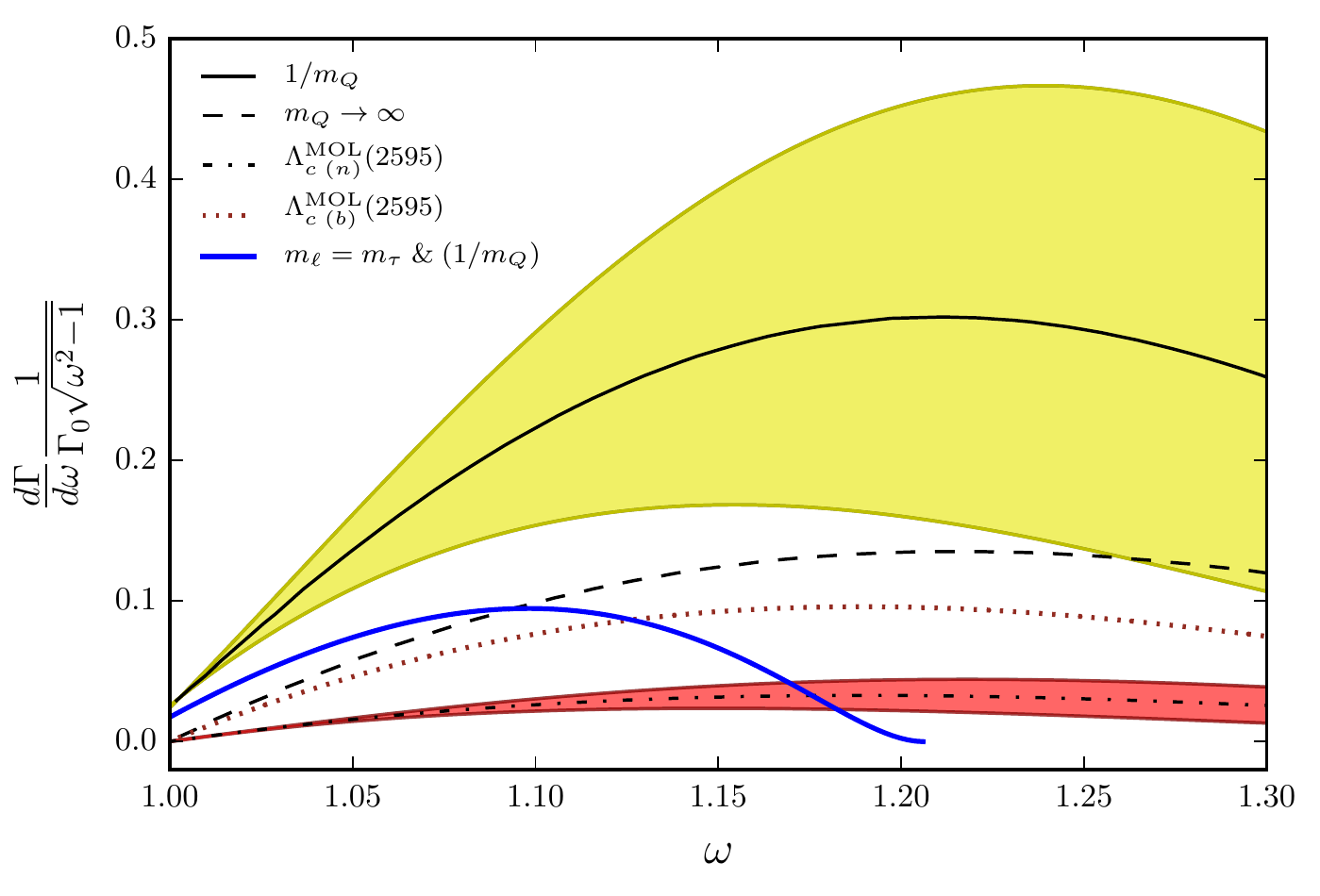} 
\caption{Differential $\Lambda_b \to \Lambda_c^*(1/2^-)\, \tau\, \bar\nu_\tau$ (solid blue line) and $\Lambda_b \to \Lambda_c^*(1/2^-)\, e\, \bar\nu_e$ rates calculated using different approaches. The black dashed and solid lines, together with the error bands of the latter, are taken from Fig.1a of Ref.~\cite{Leibovich:1997az}, where the final baryon is treated as the $J^P=1/2^-$ member of the lowest-lying $j_q^P=1^-$ HQSS doublet. The dashed line shows the $m_Q\to \infty$ prediction, Eqs.~(\ref{eq:sig-num}) and (\ref{eq:me-unomenos}), while the solid line include $1/m_Q$ effects for $\sigma_1(\omega)=0$. The bands account for the changes  in the differential decay rate when $\sigma_1(1)$ is varied in the range $[-1.2, 1.2]$ GeV.   The spectrum of the $\tau-$mode, which ends around $\omega\sim 1.2$,  is calculated using Eq.~(\ref{eq:gamma_diff}) with $\Delta_{1,2}$ given in Eq.~(\ref{eq:Deltas}), and  taking $\sigma_1(\omega)=0$. (Further details on the ${\cal O}(1/m_c)$ corrections for $\tau-$decays can be found in Table  \ref{tab:taus}). On the other hand, the lowest dotted and dashed-dotted curves, together with the error bands of the latter, stand for the decay into the broad $[\Lambda_{c\, (b)}^{\rm MOL}(2595)]$ and narrow
$[\Lambda_{c\, (n)}^{\rm MOL}(2595)]$ molecular resonances found in the ${\rm SU(6)}_{\rm lsf} \times {\rm SU(2)}_{\rm HQSS}$ model of Refs.~\cite{GarciaRecio:2008dp,Romanets:2012hm}. These differential rates
have been obtained multiplying the corresponding molecular $gG$ ratios given in Table \ref{tab:ratios} by the 
${\cal O}(1/m_Q)$ improved $\Lambda_b \to \Lambda_c^*(3/2^-[j_q^P=1^-])\, e\, \bar\nu_e$ distribution displayed in Fig.1b  of Ref.~\cite{Leibovich:1997az}. Central values have been evaluated using the black solid line of this latter figure. The bands, depicted for the decay into the narrow $\Lambda_c(2595)$ molecular state, show the impact in the spectrum of the uncertainties on the ${\cal O}(1/m_Q)$ corrections, and have been calculated using the shaded region shown in Fig.1b  of Ref.~\cite{Leibovich:1997az}.}
 \label{fig:ratios}
\end{figure}

Finally, we see that the ${\rm SU(6)}_{\rm lsf} \times {\rm SU(2)}_{\rm HQSS}$ ratios for the narrow molecular $\Lambda_c(2595)$ resonance, though small ($0.14-0.28$), are neither negligible, nor totally discarded by the available data. As we expected, they are suppressed because  within this approach this state has a large $j_q^P=0^-$ {\it ldof} component. Semileptonic decays into the broad $\Lambda_c(2595)$ resonance  are about a factor of three larger, but the $\Gamma^{1/2 (b)}_{sl,\,\pi}/ \Gamma^{3/2}_{sl,\,\pi}$  ratios are still below 1/2, the $m_Q\to \infty$ prediction, and well below the IW$_{{\cal O}(1/m_Q)}$ central values obtained in \cite{Leibovich:1997az} (see Fig.~\ref{fig:ratios}). Both sets of results point to  important  $\left(1/m_Q\right)^n$  corrections, induced by the meson-baryon interactions that generate the molecular states.  On the other hand, we do not expect large variations from the consideration of hidden strange channels as intermediate states. From the couplings 
reported in Refs.~\cite{GarciaRecio:2008dp,Romanets:2012hm}, only $\Lambda\, D_s$ and $\Lambda\, D_s^*$ might be important through their coupling to the 
narrow $\Lambda_c(2595)$ state, but the respective thresholds are located (around 3.1  and 3.2 GeV) well above the resonance position, and it is not reasonable to claim for 
large effects produced by these high energy physics contributions. Actually, we have checked that the  ratios given 
in Table~\ref{tab:ratios} for the ${\rm SU(6)}_{\rm lsf} \times {\rm SU(2)}_{\rm HQSS}$ model hardly change if the large number of coupled-channels used in Refs.~\cite{GarciaRecio:2008dp,Romanets:2012hm} is reduced only to $D^{(*)}N$ and $\pi \Sigma_c^{(*)}$.

The predictions for the  ratios in molecular schemes are very sensitive to the interference and relative weights of 
the $DN$ and $D^*N$ contributions~\cite{Liang:2016exm, Liang:2016ydj}, and thus future accurate measurements of these ratios 
will shed light on the nature of the $\Lambda_c(2595)$, allowing us to address issues as the existence of two poles or the importance of the $D^*N$ channel  in the formation of the resonance(s). Such studies will also help to understand the interplay between CQM and hadron-scattering degrees of freedom~\cite{Ortega:2009hj,  Prelovsek:2013cra, Torres:2014vna, Cincioglu:2016fkm, Albaladejo:2016ztm, Albaladejo:2018mhb} in the dynamics of the $\Lambda_c(2595)$ and $\Lambda_c(2625)$. 

Note that in other molecular schemes, like the SU(4) flavor $t$-channel exchange of vector mesons  
of Refs.~\cite{Hofmann:2005sw,Mizutani:2006vq,JimenezTejero:2009vq} or those based on 
the chiral isoscalar $\pi\Sigma^{(*)}_c$ interactions~\cite{Lutz:2003jw, Hofmann:2006qx, Lu:2014ina}, where the $D^*N$ channel is not included, the $\Gamma_{\rm sl}[\Lambda_b \to \Lambda_c(2625)]$ and $\Gamma_{\rm \pi}[\Lambda_b \to \Lambda_c(2625)]$ widths will be zero or highly suppressed. This is because  the $\pi\Sigma_c^*$ pair, that dynamically generates the $\Lambda_c(2625)$ resonance in these models,  can be only produced by going beyond the spectator approximation implicit in the mechanism of Fig.~\ref{fig:SL}. This places an additional limitation on the validity of these approaches, which already have some problems to describe the mass and width of the $\Lambda_c(2625)$ (see the related discussion  in Subsecs.~\ref{sec:su4} and \ref{sec:chiral}).

\subsection{$\Lambda_b \to \Lambda_c(2595)\tau\,\bar\nu_\tau$ and $\Lambda_b \to \Lambda_c(2625)\tau\,\bar\nu_\tau$ decays}
\label{sec:results-tau}
Let us now pay attention to $\Lambda_b$ semileptonic decays with a $\tau$ lepton in the final state. At the LHC, a large number of ground-state $\Lambda_b$ baryons are 
produced~\cite{Aaij:2011jp}, and its decays into charmed baryons can be used to constrain violations of  LFU. These decays are of interest in light
of the $R(D^{(*)})$ puzzle in the semileptonic $\bar B \to D^{(*)}\tau\,\bar\nu_\tau$ decays (see for instance the discussion in~\cite{Cerri:2018ypt}, and references therein). 
Decays involving the ground state charmed baryon, $\Lambda_c$,  have been already studied in lattice QCD~\cite{Detmold:2015aaa} 
and beyond the Standard Model~\cite{Datta:2017aue}.  On the other hand, the LHCb collaboration has reported large samples of $\Lambda_c(2595)$ 
and $\Lambda_c(2625)$ baryons in $\Lambda_b$ semileptonic decays~\cite{Aaij:2017svr}, which makes meaningful to investigate the LFU ratios~\cite{Boer:2018vpx, Gutsche:2018nks} 
\begin{equation}
 R[\Lambda_c^*] = \frac{{\cal B}(\Lambda_b \to \Lambda_c^* \tau\,\bar\nu_\tau)}{{\cal B}(\Lambda_b \to \Lambda_c^* \mu\,\bar\nu_\mu)}
\end{equation}
with $\Lambda_c^*= \Lambda_c(2595)$ or $\Lambda_c(2625)$, due to the good prospects that LHCb can measure them in the short term. Results are shown in 
Table~\ref{tab:taus}.

We have used Eq.~(\ref{eq:gamma_diff}) to compute 
$\Gamma[\Lambda_b \to \Lambda_c(2595)\tau\,\bar\nu_\tau]$ and $\Gamma[\Lambda_b \to \Lambda_c(2625)\tau\,\bar\nu_\tau]$,  assuming that 
the $\Lambda_c(2595)$ and $\Lambda_c(2625)$  form the lowest-lying   $j_q^P=1^-$ HQSS doublet, and have taken the ${\cal O}(1/m_c)$ improved form factors given in Eq.~(\ref{eq:betas}). Therefore, spin symmetry in the $b-$quark sector is conserved,   which implies neglecting terms of order $\Lambda_{\rm QCD}/m_b$. This is an excellent approximation, and we reproduce within a 5\% the $\Lambda_c(2595)$ differential and integrated rates reported in Ref.~\cite{Leibovich:1997az}. The approximation works even better for the $\Lambda_c(2625)$,  and  moreover it leads to simple expressions for the $\omega-$differential widths, including full finite-lepton mass contributions that
are necessary for testing LFU. Note that the calculations of Ref.~\cite{Leibovich:1997az} were made in the $m_\ell \to 0$ limit.  

Predictions for semileptonic $\tau-$decays are relatively stable against the uncertainties on the ${\cal O}(1/m_c)$ corrections, 
because in this case $\omega_{\rm max}\sim 1.2$, and the largest contributions to the integrated width come from regions relatively  close to zero recoil (see blue solid line of Fig.~\ref{fig:ratios}). However, there are still some uncertainties associated with the lack of information about the form factor $\sigma_1(\omega) $, although they are significantly smaller than those shown in Table \ref{tab:ratios} for the case of massless leptons. The $\sigma_1$ term produces, also for $\tau-$decays, opposite effects for $\Lambda_c(2595)$ or $\Lambda_c(2625)$ final states (see Eqs.~(\ref{eq:betas}) and (\ref{eq:cjs})). Uncertainties partially cancel in the  $R[\Lambda_c(2595)]$ and $R[\Lambda_c(2625)]$ ratios, which are predicted in Table \ref{tab:taus}  with moderate errors. We expect these ratios to be comprised in the intervals [0.10, 0.15] and [0.10, 0.13], respectively. {These estimates compare rather well with those obtained in the covariant confined quark model employed in Ref.~\cite{Gutsche:2018nks}.} 

Next we discuss the $\Gamma^{1/2}_{\rm sl;\, \tau}/ \Gamma^{3/2}_{\rm sl;\, \tau}$ ratio, for which theoretical errors  are larger. The  central value of this ratio compares rather well with that quoted in Table \ref{tab:ratios} for light leptons ($\mu$ or $e$), though its errors for the $\tau$ mode are  slightly smaller. 

The $\Gamma^{1/2}_{\rm sl;\, \tau}/ \Gamma^{3/2}_{\rm sl;\, \tau}$ ratio would drastically change if the final charmed baryons turned out to be  predominantly hadronic molecules. In that situation, we would obtain the same values as in Table \ref{tab:ratios} from the $gG$ factors compiled in  Table~\ref{tab:gG}. We should point out that because the available phase space is smaller for the $ \tau $ mode, the decay most likely occurs near the zero-recoil point where the approximations that lead to the quotient of $gG$ factors in Eq.~(\ref{eq:decaymol}) are more precise. The predicted ratios would depend on the molecular scheme, and on the  member of the double pole structure of the $\Lambda_c(2595)$ involved in the decay. However, in all cases, we would obtain values  below 0.5, at least one-sigma away from the predictions collected in Table \ref{tab:taus},  based on the hypothesis that the $\Lambda_c(2595)$ 
and $\Lambda_c(2595)$ form the lowest-lying $1^-$ HQSS multiplet of excited charm-baryons. This latter picture also discards the existence of a second $J^P=1/2^-$ (broad) resonance  in the 2.6 GeV region. 

It is not clear how the $R[\Lambda_c(2595)]$  and $R[\Lambda_c(2625)]$ ratios would be affected if any of the resonances has a large molecular component, since this will also affect the decay widths into light leptons that appear in the denominators of these ratios. Therefore, one might think that they would not be significantly modified  with respect to the values given in Table \ref{tab:taus}, that mostly account for the reduction of phase space. Nevertheless, it is difficult to be more quantitative. However, $R[\Lambda_c(2595)]$ may be affected by 
a new source of potentially large systematic errors, if in the $\tau$ and  $\mu$ or $e$ modes, the same $\Lambda_c(2595)$ molecular state is not observed. This confusion  would produce large numerical variations that would suggest false violations of LFU.

\begin{table*}[t!]
\begin{tabular}{cc|cc}
$\Gamma[\Lambda_b \to \Lambda_c(2595)\tau\,\bar\nu_\tau$] & $\Gamma[\Lambda_b \to \Lambda_c(2595)\mu\,\bar\nu_\mu$] & $\Gamma[\Lambda_b \to \Lambda_c(2625)\tau\,\bar\nu_\tau]$ & $\Gamma[\Lambda_b \to \Lambda_c(2625)\mu\,\bar\nu_\mu]$  \\\hline
$0.55_{-0.18}^{+0.23}(\sigma_1)^{+0.19}_{-0.15} (\bar\Lambda')$ & $4.8\pm 2.4 (\sigma_1)^{+1.3}_{-1.1} (\bar\Lambda')$ & $0.38_{-0.08}^{+0.09}(\sigma_1)$ &$3.5_{-1.2}^{+1.3}(\sigma_1)$ \\\hline\hline
\end{tabular}
\begin{tabular}[t]{ccc|cc}
 $R[\Lambda_c(2595)]$ & $R[\Lambda_c(2625)]$ & $\Gamma^{1/2}_{\rm sl;\, \tau}/ \Gamma^{3/2}_{\rm sl;\, \tau}$& $R[\Lambda_c(2595)]$~\cite{Gutsche:2018nks} & $R[\Lambda_c(2625)]$~\cite{Gutsche:2018nks} \\\hline
 $0.11_{-0.01}^{+0.04}$ &$0.11_{-0.01}^{+0.02}$ &$1.5^{+1.2}_{-0.8}$ &$0.13\pm0.03$ &$0.11\pm 0.02$\\\hline
\end{tabular}
 \caption{Semileptonic decay widths $\Gamma[\Lambda_b \to \Lambda_c^*\,\ell\,\bar\nu_\ell$] (in $\Gamma_0/100$ units) for $\mu$ and $\tau$ modes. The  rates are calculated using Eq.~(\ref{eq:gamma_diff}), with form-factors given in Eqs.~(\ref{eq:Deltas}) and (\ref{eq:Omegas}) or equivalently in Eq.~(\ref{eq:betas}). They contain the subleading ${\cal O}(1/m_c)$ corrections derived  in  Ref.~\cite{Leibovich:1997az}, assuming that the $\Lambda_c(2595)$ and $\Lambda_c(2625)$  form the lowest-lying   $j_q^P=1^-$ HQSS doublet. We also show the $\tau/\mu-$semileptonic ratios for both final baryon states, and $\Gamma_{\rm sl}[\Lambda_b \to \Lambda_{c}(2595)]/\Gamma_{\rm sl}[\Lambda_b \to \Lambda_c(2625)]$ for the $\tau\,\bar\nu_\tau$ semileptonic mode. Errors are derived from the uncertainties on the $\sigma_1$ form-factor and the $\bar\Lambda'$ parameter, and are added in quadrature for the ratios shown in the last three columns. Central values are obtained for $\sigma_1(\omega)=0$ in all cases.
 {Results from Ref.~\cite{Gutsche:2018nks} are taken from TABLE~II.}} \label{tab:taus} 
\end{table*}

{Finally, in Table~\ref{tab:taugsLc} we collect  several predictions~\cite{Shivashankara:2015cta,Li:2016pdv,Faustov:2016pal,Azizi:2018axf,DiSalvo:2018ngq,Gutsche:2018nks} of the  LFU ratios for the $\Lambda_b$ semileptonic decay into the ground-state $\Lambda_c$ $(1/2^+)$.
\begin{table}[t!]
    \centering
    \begin{tabular}{c|c|c|c|c|c}
    $R[\Lambda_c]$~\cite{Shivashankara:2015cta}&$R[\Lambda_c]$~\cite{Li:2016pdv}&$R[\Lambda_c]$~\cite{Faustov:2016pal}&$R[\Lambda_c]$~\cite{Azizi:2018axf}&$R[\Lambda_c]$~\cite{DiSalvo:2018ngq}&$R[\Lambda_c]$~\cite{Gutsche:2018nks}\\\hline
    $0.29\pm 0.02$&$0.26\sim 0.34$ &$0.313$ &$0.31\pm 0.11$ &$0.15\sim 0.18$ &$0.30\pm 0.06$ \\\hline
    \end{tabular}
    \caption{{Semileptonic $\Lambda_b \to \Lambda_c (1/2^+)$
    LFU ratios, $R[\Lambda_c]$, obtained in the works of Refs.~\cite{Shivashankara:2015cta,Li:2016pdv,Faustov:2016pal,Azizi:2018axf,DiSalvo:2018ngq,Gutsche:2018nks}.}}
    \label{tab:taugsLc}
\end{table}
Comparing the ratios of Tables~\ref{tab:taus} and \ref{tab:taugsLc}, we see that  $R[\Lambda_c]$ is predicted to be significantly larger than $R[\Lambda_c^*]$. (Note, however, that the result of Ref.~\cite{DiSalvo:2018ngq} is considerably smaller than those given by the other authors.)}

%
\section{Summary}
\label{sec:summary}
In this work, we have studied the  $\Lambda_b \to  \Lambda_c^*\ell\bar{\nu}_\ell$ and  $\Lambda_b \to \Lambda_c^*\pi^-$ $[\Lambda_c^*=\Lambda_c(2595)$ and 
$\Lambda_c(2625)]$ decays, paying special attention to the implications that can be derived from HQSS. We have critically reviewed different 
molecular descriptions of these charm excited baryons, and have discussed in detail the main features of those schemes that predict a two-pole  pattern for the $\Lambda_c(2595)$, in analogy to the case of the similar $\Lambda(1405)$ resonance in the strange sector. 

We have calculated the ratios  $\Gamma(\Lambda_b\rightarrow\Lambda_c(2595)\pi^-)/\Gamma(\Lambda_b\rightarrow\Lambda_c(2625)\pi^-)$
  and $\Gamma(\Lambda_b\rightarrow \Lambda_c(2595)\, \ell
  \,\bar{\nu}_\ell)/ \Gamma(\Lambda_b\rightarrow\Lambda_c(2625)\, \ell
  \,\bar{\nu}_\ell)$, and have shown that molecular schemes are very sensitive to the interference and relative weights of 
the $DN$ and $D^*N$ contributions, as firstly pointed out in Refs.~\cite{Liang:2016exm, Liang:2016ydj}. Actually, we have re-derived some of the results of these latter works using  a manifest Lorentz and HQSS invariant formalism. In this context, we have argued that future accurate measurements of the above ratios will shed light on the nature of the $\Lambda_c(2595)$, allowing us to address issues as the existence of two poles or the importance of the $D^*N$ channel  in the formation of the resonance(s). 

We have also  investigated the LFU ratios $R[\Lambda_c^*] = {\cal B}(\Lambda_b \to \Lambda_c^* \tau\,\bar\nu_\tau)/{\cal B}(\Lambda_b \to \Lambda_c^* \mu\,\bar\nu_\mu)$. We have  computed  $\Gamma[\Lambda_b \to \Lambda_c(2595)\tau\,\bar\nu_\tau]$ and $\Gamma[\Lambda_b \to \Lambda_c(2625)\tau\,\bar\nu_\tau]$  assuming that 
the $\Lambda_c(2595)$ and $\Lambda_c(2625)$  form the lowest-lying   $j_q^P=1^-$ HQSS doublet, and have taken ${\cal O}(1/m_c)$ improved form factors~\cite{Leibovich:1997az}.  We have used a scheme that preserves spin-symmetry in  the $b-$quark sector, which implies neglecting corrections of order $\Lambda_{\rm QCD}/m_b$. This is an excellent approximation that leads to simple expressions for the $\omega-$differential widths, including full finite-lepton mass contributions that
are necessary for testing LFU.

Finally, we have pointed out that the $R[\Lambda_c(2595)]$ ratio may be affected by  a new source of potentially large systematic 
errors if there are two  $\Lambda_c(2595)$ poles. 

At the LHC, a large number $\Lambda_b$ baryons are 
produced, and the LHCb collaboration has reported large samples of $\Lambda_c(2595)$ 
and $\Lambda_c(2625)$ baryons in its semileptonic decays. Hence, there are good prospects that LHCb can measure in the near future some of the ratios discussed in this work.

\section*{Acknowledgments}
We warmly thank E. Oset for useful discussions. R.P.~Pavao  wishes to thank the program Santiago Grisolia of the Generalitat Valenciana. This research  has been supported by  
 the Spanish Ministerio de Ciencia, Innovaci\'on  y Universidades and European FEDER funds under  Contracts FIS2017-84038-C2-1-P
 and  SEV-2014-0398.
 S.~Sakai acknowledges the support by NSFC and DFG through funds provided to the Sino-German CRC110 ``Symmetries and the Emergence of Structure in QCD'' (NSFC Grant No.~11621131001), by the NSFC (Grant No.~11747601 and No.~11835015), by the CAS Key Research Program of Frontier Sciences (Grant No.~QYZDB-SSW-SYS013), by the CAS Key Research Program (Grant No.~XDPB09) and by the CAS President’s International Fellowship   Initiative (PIFI) (No. 2019PM0108).

\bibliography{biblio}
\end{document}